\documentclass[aps,preprint,nofootinbib]{revtex4}%
\usepackage{slashed}
\usepackage{url,hyperref}
\usepackage{amsfonts}
\usepackage{amsmath}
\usepackage{amssymb}
\usepackage{graphicx}%
\providecommand{\U}[1]{\protect\rule{.1in}{.1in}}

\begin{document}
\title{Locally Scale Invariant Chern-Simons Actions in 3+1 Dimensions and Their
Emergence From 4+2 Dimensional 2T-Physics}
\author{Itzhak Bars}
\email{bars@usc.edu}
\affiliation{Department of Physics, University of Southern California, Los Angeles, CA 90089}
\author{Sophia D. Singh}
\email{sophia\_singh@brown.edu}
\affiliation{Department of Physics, Brown University, Providence, RI 02912}

\begin{abstract}
The traditional Chern-Simons (CS) terms in 3+1 dimensions that modify General
Relativity (GR), Quantum Chromodynamics (QCD), and Quantum Electrodynamics
(QED), typically lack scale invariance. However, a locally scale invariant and
geodesically complete framework for the Standard Model (SM) coupled to GR was
previously constructed by employing a tailored form of local scale (Weyl)
symmetry. This refined SM+GR model closely resembles the conventional SM in
subatomic realms where gravitational effects are negligible. Nevertheless, it
offers an intriguing prediction: the emergence of new physics beyond the
traditional SM and GR near spacetime singularities, characterized by intense
gravity and substantial deviations in the Higgs field. In this study, we
expand upon the enhanced SM+GR by incorporating Weyl invariant CS terms for
gravity, QCD, and QED in 3+1 dimensions, thereby integrating CS contributions
within the locally scale-invariant and geodesically complete paradigm.
Additionally, we establish a holographic correspondence between the new CS
terms in 3+1 dimensions and novel 4+2 dimensional CS-type actions within
2T-physics. We demonstrate that the Weyl transformation in 3+1 dimensions
arises from 4+2 general coordinate transformations, which unify the hidden
extra 1+1 large (not curled up) dimensions with the evident 3+1 dimensions. By
leveraging the newfound local conformal symmetry, the augmented and
geodesically complete SM+GR+CS introduces innovative tools and perspectives
for exploring classical field theory aspects of black hole and cosmological
singularities in 3+1 dimensions, while the 4+2 dimensional connection unveils
deeper facets of spacetime.

\end{abstract}
\maketitle
\tableofcontents

\newpage

\section{Introduction}

Chern-Simons (CS) terms that modify General Relativity (GR) \cite{Jackiw} find
their theoretical underpinnings in both field theory
\cite{Anomalies,Axial,Bell} and string theory \cite{loop}. In the realm of
field theory, the emergence of the gravitational CS term stems from the stress
tensor trace anomaly, while in particle physics, it arises from the chiral
anomaly involving axial currents \cite{Axial, Bell}. In string theory, the CS
term serves to rectify anomalous symmetries \cite{loop}. Thus, the prevalence
of CS terms in GR, QCD, and QED is well-founded from a theoretical standpoint.

Initially conceived as a 2+1 dimensional topological theory \cite{Jackiw2},
Chern-Simons gravity later found extension to 3+1 dimensions by embedding the
3-dimensional Chern-Simons topological current into a 4-dimensional spacetime
manifold. The conventional CS term within the 3+1 dimensional gravity action
as formulated by \cite{Jackiw} is represented by:
\begin{equation}
S_{\text{$CS-GR$}}^{3+1}=\int d^{4}x~\tilde{R}R\times\left(  \text{scalar
field}\right)  ,\;\;\;\tilde{R}R\equiv\frac{1}{2}\epsilon^{\mu_{1}\mu_{2}%
\mu_{3}\mu_{4}}R_{\;\sigma\mu_{1}\mu_{2}}^{\lambda}R_{\;\lambda\mu_{3}\mu_{4}%
}^{\sigma}. \label{RR*}%
\end{equation}
This term supplements the standard Einstein-Hilbert term, kinetic terms for
the scalar field, and other matter fields and their interactions.
Consequently, the introduction of the GR field equations is accompanied by
modifications through the inclusion of an additional Cotton-like C-tensor,
constructed from derivatives of the Ricci tensor and the dual of the Riemann
tensor. This C-tensor emerges when the Chern-Simons action $S_{\text{$CS-GR$}%
}^{3+1}${} is varied with respect to its spacetime metric $g_{\mu\nu}$.
Similar modifications are evident in QCD when the curvature tensor
$R_{~\sigma\mu\nu}^{\lambda}$ is replaced by the Yang-Mills field strength
tensor $F_{\mu\nu}^{a}$ and similarly in QED when $R_{~\sigma\mu\nu}^{\lambda
}$ is replaced by the electromagnetic field strength tensor $F_{\mu\nu}$.

The CS corrections to GR, QCD, and QED yield significant physical
implications. For instance, CS-modified gravity \cite{Birefringent}%
-\cite{Spin} has been instrumental in elucidating inflationary leptogenesis
and baryogenesis through the gravitational anomaly. An intriguing aspect of
this mechanism is the prediction of amplitude birefringent gravitational
waves, potentially leaving a distinctive imprint on the gravitational wave
background and thus on the Cosmic Microwave Background. Investigating
Chern-Simons gravity could prove crucial in addressing inquiries regarding the
evolutionary trajectory of the universe. Similarly, in QCD the postulation of
an axion field aims to circumvent the breakdown of CP symmetry in strong
interactions \cite{PQ}, while in QED the decay of the pion into two photons,
among other phenomena, is encapsulated by a CS term within an effective
action. Some additional discussions of the CS terms include \cite{witten}%
-\cite{kimura}.

\subsection{Geodesically complete spacetime \label{incompleteST}}

This paper delves into how Chern-Simons (CS) terms influence physics in
regions of intense gravity. However, before discussing this topic, it is
essential to recognize that our conventional tools in modern physics have been
insufficient in describing gravity near black hole (BH) and cosmological
singularities. Beyond the anticipated but not fully comprehended quantum
gravity effects, there lies the issue of spacetime's geodesic incompleteness,
which arises in the Standard Model coupled to General Relativity. It appears
that this incompleteness has not garnered the warranted attention, perhaps due
to the anticipation that quantum gravity would resolve singularity problems.
Nonetheless, the tools employed to explore potential quantum gravity theories,
such as string theory, also exhibit geodesic incompleteness.\footnote{To
illustrate, gravitational background fields in particle or string theory
exhibit geodesic incompleteness akin to their field theory counterparts. The
presence of a dilaton or other background fields fails to rectify this issue.
Similar to particle geodesics, strings propagating in such backgrounds
encounter singularities of the background gravitational metric within a finite
amount of proper time. What happens beyond this duration remains elusive as
the standard theory offers no insights. This illustrates the problem of
geodesic incompleteness in particle theory, field theory, and string theory at
the classical level. There is a conspicuous absence of enlightening
discussions on how quantum string theory or other quantum gravity theories
might resolve this issue. However, refer to \cite{BSTstrings} for a proposed
modification of string theory which is linked to the geodesic completion
mechanism discussed here.}

This highlights that resolving geodesic incompleteness, as outlined below,
would inherently furnish new methodologies for investigating hitherto unknown
physical phenomena within black holes and in the vicinity of the early
universe, across both classical and quantum domains.

A geodesically complete framework for the Standard Model coupled to General
Relativity (SM+GR) in 3+1 dimensions naturally emerged from 2T-physics in 4+2
dimensions \cite{Survey}-\cite{book}. The higher-dimensional theory, denoted
as (SM+GR)$_{4+2}$ (refer to section \ref{sec:2T-physics-approach} and
appendix \ref{A1}), has a high degree of gauge symmetry which, in a
gauge-fixed rendition, yields a holographic image in 3+1 dimensions that
encapsulates all the gauge-invariant information of the 4+2 formalism. Termed
the \textquotedblleft holographic conformal shadow" (see section
\ref{HolShad}), this image serves as the foundation for the conformally
improved Standard Model coupled to gravity \cite{BST}, denoted herein as
i(SM+GR)$_{3+1}$ (refer to section \ref{sec:building-weyl-symmetric}).
Geodesic completion emerged as an incidental property of the holographic
conformal shadow, albeit not the primary objective of the 2T-physics
formalism. Even without delving into the details of 2T-physics, one can
outline the critical attributes of the enhanced i(SM+GR)$_{3+1}$, leading to
geodesic completeness and foreseeing hitherto unimaginable physics in regions
of intense gravity, particularly in the proximity of gravitational
singularities \cite{BbBc}\cite{BHoles}\cite{BH+Higgs}.

There are two primary components for geodesic completeness:

\begin{itemize}
\item The first involves a local scale (Weyl) symmetry which is inherently
mandated by any relativistic field theory coupled to gravity in 3+1
dimensions, provided it is derived from 2T-physics. This local scale symmetry
arises from residual effects of general coordinate transformations in 4+2
dimensions that mix the hidden extra 1+1 dimensions with the evident 3+1
dimensions (refer to section \ref{origins}). Scale invariance prohibits
dimensionful parameters, implying that if 2T-physics is the viable approach,
then all dimensionful parameters in Nature must originate from gauge fixing
and/or spontaneous breaking of the emergent Weyl symmetry. In i(SM+GR)$_{3+1}%
$, a single real field $\phi\left(  x\right)  $ serves as the sole source,
generating all dimensionful parameters, including the gravitational Newton
constant $G_{N}$, the cosmological constant $\Lambda$, the Higgs mass or
equivalently its vacuum value $v$ (VEV), and the masses of quarks, leptons,
and gauge bosons that are related to the Higgs field (refer to section
\ref{DimParm}).

\item While the local scale invariance in i(SM+GR)$_{3+1}$ has a distinct
origin from Weyl's original concept (see section \ref{origins}), they may
appear indistinguishable from the perspective of 3+1 dimensions. However, in
the enhanced i(SM+GR)$_{3+1}$, it manifests with a uniquely specialized
structure. This insight emerged in 2008 with the introduction of Gravity in
2T-physics \cite{2TgravIB}\cite{2Tgrav}, and was further elaborated upon
directly in 3+1 dimensions \cite{BST}. This structure entails a coordinated
interplay between the singlet field $\phi\left(  x\right)  $ and the Higgs
doublet $H(x)$. Both must serve as conformally coupled scalars to maintain
local scale invariance. As a consequence, instead of the traditional SM+GR's
Einstein-Hilbert term $\left(  16\pi G_{N}\right)  ^{-1}R\left(  g\right)  $,
this must be adjusted to $\frac{1}{12}\left(  \phi^{2}-2H^{\dagger}H\right)
R\left(  g\right)  $ in the enhanced i(SM+GR)$_{3+1}.$ The scalars $\phi,H$
must possess opposite signs in their non-minimal coupling to the curvature $R$
to establish the existence of a region of spacetime where the resulting
dynamical gravitational strength $G\left(  x\right)  $ is positive while
requiring the Higgs field to have the correct sign kinetic energy term in the
action. Then $\phi$ must have the wrong sign kinetic term, but this ghost-like
$\phi$ causes no issues with unitarity thanks to the local gauge symmetry that
compensates for the ghost-like unphysical gauge degree of freedom (refer to
section \ref{sec:building-weyl-symmetric}). This distinctive structure with a
relative minus sign for conformally coupled scalars could theoretically have
been recognized directly in 3+1 dimensions prior to 2T-physics, but it
remained overlooked.
\end{itemize}

Together, these two components imply that the strength of gravity, denoted as
$G\left(  x\right)  $, is not uniform across the universe. Instead, the
strength of gravity varies dynamically with spacetime as determined by the
scalar fields $\phi$ and $H$ and is given by
\begin{equation}
\left(  16\pi G\left(  x\right)  \right)  ^{-1}=\frac{1}{12}\left(  \phi
^{2}\left(  x\right)  -2H^{\dagger}H\left(  x\right)  \right)  .
\label{Newton}%
\end{equation}

In the context of i(SM+GR)$_{3+1}$, the familiar low-energy physics operates
within a spacetime region where the local scale-invariant ratio $\left(
2H^{\dagger}H\right)  /\phi^{2}$ is negligible. In this regime, $\phi$ is of
the order of the Planck scale $10^{19}$ GeV, and the Higgs field $H$ is around
246 GeV, including its vacuum expectation value (VEV). To study physics in
this low-energy regime, one can adopt a gauge where $\phi\left(  x\right)  $
is constant, $\phi\left(  x\right)  =\phi_{0}\sim$ $10^{19}$ GeV, treating the
Higgs field $H\sim$ 246 GeV as negligible at low energies in the expression
for the almost constant gravitational strength $G\left(  x\right)  \simeq
G_{N}${} in Eq.(\ref{Newton}). In this gauge, only the Higgs field remains as
a physical spin-0 degree of freedom while all dimensional physical constants
emerge as proportional to $\phi_{0}${} (see section \ref{DimParm}). This
elucidates how and why i(SM+GR)$_{3+1}$ closely resembles the traditional
SM+GR in all observed low-energy physics aspects. Thus, i(SM+GR)$_{3+1}${}
stands as a valid theory for all experimentally observed low-energy physics to
date, instilling confidence in its predictions outlined below in obscure or
unknown spacetime regions or energy scales.

It is of interest to contrast the physical predictions of i(SM+GR)$_{3+1}$
versus traditional SM+GR in domains beyond low energy physics where the theory
is uncertain as follows.

$\left(  i\right)  $ The field equations derived from i(SM+GR)$_{3+1}$
indicate that near gravitational singularities, the Weyl gauge-invariant ratio
$\left(  2H^{\dagger}H\right)  /\phi^{2}$ approaches 1, suggesting that
$\left(  \phi^{2}-2H^{\dagger}H\right)  $ can vanish, resulting in a divergent
effective gravitational strength $G\left(  x\right)  $ precisely at the
singularity. On one side of the singularity, where the gauge-invariant ratio
is less than one, $\left(  2H^{\dagger}H\right)  /\phi^{2}<1$, gravity is
attractive $G\left(  x\right)  >0$; on the other side, where the ratio exceeds
one, $\left(  2H^{\dagger}H\right)  /\phi^{2}>1$, gravity is repulsive
$G\left(  x\right)  <0$. Consequently, i(SM+GR)$_{3+1}${} predicts
geodesically complete spacetime configurations where gravitational
singularities separate gravity regions from antigravity regions. By contrast,
traditional SM+GR only encompasses the gravity side of the singularity,
contributing to its geodesic incompleteness.\footnote{The geodesic
incompleteness in SM+GR becomes apparent when considering its accommodation
within (SM+GR)$_{3+1}$, albeit with incomplete gauge choices. An example of a
geodesically incomplete gauge is the E-gauge \cite{BST} where fields $\left(
\phi_{E},H_{E},g_{E\mu\nu}\right)  $ are denoted by the letter `E' to
differentiate them from other gauges. In this gauge, the fields obey $\frac
{1}{12}\left(  \phi_{E}^{2}\left(  x\right)  -2H_{E}^{\dagger}\left(
x\right)  H_{E}\left(  x\right)  \right)  =\left(  16\pi G_{N}\right)  ^{-1}$
to reproduce the Einstein frame featuring the standard Einstein-Hilbert term
$\left(  16\pi G_{N}\right)  ^{-1}R\left(  g_{E}\right)  $, with a
spacetime-independent positive Newton constant $G_{N}$. However, the E-gauge
exhibits geodesic incompleteness, since it is valid only in spacetime patches
where the gauge-invariant expression $\left(  1-2H^{\dagger}H/\phi^{2}\right)
${} is positive. Notably, this expression's sign cannot be altered by local
scale transformations, leading to regions where the Weyl invariant expression
$\left(  1-2H^{\dagger}H/\phi^{2}\right)  $ is negative being omitted.
Furthermore, the E-gauge fails to describe events at spacetime singularities
where $\left(  1-2H^{\dagger}H/\phi^{2}\right)  =0$. Conversely, it has been
demonstrated \cite{Cyclic} that in other gauges, the generic solutions of the
equations of motion for $\phi$ and $H$ continuously span all signs of the
gauge invariant $\left(  1-2H^{\dagger}H/\phi^{2}\right)  $, encompassing both
gravity and antigravity patches. This underscores the evident geodesic
incompleteness of the traditional SM+GR framework in the Einstein frame.
Similarly, any Jordan frame where the effective gravitational strength
$G\left(  x\right)  $ is positive exhibits geodesic incompleteness. This
incompleteness extends to the string frame that emerges in the low-energy
limit of string theory, as it too constitutes a Jordan frame with solely
positive effective gravitational strength $G\left(  x\right)  $. While such
incomplete frames are attainable via Jordan-type gauge fixing in
(SM+GR)$_{3+1}$ \cite{BST}\cite{BSTstrings}, their incompleteness issue is
rectified by incorporating the antigravity regions beyond singularities
predicted in (SM+GR)$_{3+1}$ as outlined in Section
\ref{sec:building-weyl-symmetric}. \label{EJsframes}}

$\left(  ii\right)  $ In i(SM+GR)$_{3+1}$, it has been established that the
majority of generic classical cosmological field solutions \cite{Cyclic}
describe fields propagating analytically through singularities from gravity
regions to antigravity regions and vice versa. These typical solutions
encompass the majority of phase space for on-shell cosmological field
configurations \cite{Cyclic}. The divergence of $G\left(  x\right)  $ at
spacetime regions or points where $\left(  2H^{\dagger}H\right)  /\phi^{2}=1$
does not hinder the continuity of fields or information flow through
gravitational singularities. Similarly, geodesics connecting gravity and
antigravity regions remain continuous at the singularity. Local scale
invariance significantly contributes to establishing the continuity of
information flow across the singularity \cite{Cosmo2}-\cite{Sailing}.

Thus, the gravity and antigravity regions connected at gravitational
singularities represent geodesic completions of each other. These prevalent
features in classical field solutions of the geodesically complete
i(SM+GR)$_{3+1}${} signify a paradigm shift in understanding physical
phenomena and information flow at singularities from one side to the other.

Conversely, the geodesic incompleteness of the traditional SM+GR framework in
the Einstein or Jordan frames, including the String frame, leads to the
inability to explore physics close to singularities at the classical field
theory level.

$\left(  iii\right)  $ Assuming that the geodesically complete field solutions
in i(SM+GR)$_{3+1}$ (similar to those discussed in \cite{Cyclic}) dominate the
phase space, a semi-classical path integral or WKB approach provides a fresh
perspective on the flow and conservation of quantum probability across both
gravity and antigravity sides of gravitational singularities \cite{Cosmo2}.
This necessitates including observers on both sides of the singularities.
Consequently, the information loss problem in black holes undergoes a radical
transformation both technically and conceptually. It becomes evident that
information lost by observers on one side of the singularity is gained by
observers on the opposite side. By incorporating all observers, unitarity
remains continuously preserved as the black hole evolves, whether it
evaporates or not.

Some analysis of the features of i(SM+GR)$_{3+1}$ described above is evident
in applications to the Big Bang \cite{BbBc}\cite{Sailing}\cite{Cosmo2}, black
holes \cite{IB+Araya}, and the interpretation of the antigravity regime
\cite{IBalbin}. Additionally, a recent discovery of intriguing behavior of the
Higgs field inside black holes will be discussed in \cite{BH+Higgs}%
.\footnote{Similar effects can be anticipated within the conventional
framework of supergravity (SUGRA), as it presents the potential for geodesic
completeness---a feature previously overlooked before the advent of
2T-physics: the curvature term in SUGRA is given by $\left(  \left(  16\pi
G_{N}\right)  ^{-1}-\frac{1}{6}K\left(  \varphi,\bar{\varphi}\right)  \right)
R\left(  g\right)  $ where $K\left(  \varphi,\bar{\varphi}\right)  $ is the
K\"{a}hler potential. Hence, SUGRA exhibits a sign-changing gravitational
strength $G\left(  x\right)  $ similar to Eq.(\ref{Newton}) when $\phi\left(
x\right)  \rightarrow\phi_{0}$. Previous SUGRA literature \cite{Weinberg}
fixated solely on the positive $G(x)$ spacetime patch, and constrained it to
remain positive through a Weyl transformation to the Einstein frame,
inadvertently resulting in geodesic incompleteness. SUGRA was elevated to a
Weyl-symmetric version that incorporates a complex superfield version of
$\phi\left(  x\right)  $ \cite{sugra4+2}\cite{BST}. Hence, SUGRA is
geodesically complete, akin to i(SM+GR)$_{3+1},$ provided the presence of the
antigravity regions that complete the spacetime are acknowledged.}

As mentioned, i(SM+GR)$_{3+1}${} represents the holographic conformal shadow
of its 2T-physics counterpart, (SM+GR)$_{4+2}$. The latter incorporates gauge
degrees of freedom that reveal the underlying 4+2-dimensional spacetime, akin
to revealing the SO$(3,1)$ Lorentz symmetry of electromagnetism and its
underlying relativistic spacetime through the inclusion of gauge degrees of
freedom in the gauge potential $A_{\mu}$. Similarly, the gauge degrees of
freedom in 2T-physics unveil spacetime symmetries such as SO$(4,2)$ in the
flat limit and its extension to coordinate invariance in curved
4+2-dimensional spacetime. These spacetime symmetries remain gauge invariant
under the Sp$(2,R)$ gauge symmetry underlying 2T-physics (See appendix
\ref{A1}). Therefore, even after gauge fixing to 3+1 dimensions, the 4+2
spacetime symmetries persist as properties of the physical systems emerging in
all shadows of 2T-physics. For instance, the familiar SO$(4,2)$ conformal
symmetry in 3+1 dimensions and the hidden SO$(4,2)$ symmetry of planetary
motion or the Hydrogen atom, among other cases, are elucidated by the evident
4+2-dimensional Lorentz symmetry in 2T-physics. Particularly, the Weyl
symmetry in i(SM+GR)$_{3+1}${} is a remnant of general coordinate
reparametrizations in 4+2 dimensions that has not yet been gauge-fixed in the
holographic conformal shadow (discussed in section \ref{origins}).

Since 2T-physics underpins i(SM+GR)$_{3+1}$, it is pertinent to briefly
describe its origins and the theory's relationship with 1T-physics. The
following provides essential insights into the concepts of 2T-physics. Further
details can be found as a concise summary in appendix \ref{A1}.

At the core of 2T-physics there is a gauge symmetry in phase space
$(X^{M},P_{M})$, postulating that all fundamental physical laws must treat
position and momentum equally. This generalization extends Einstein's general
coordinate invariance to phase space rather than just position space. This
phase space gauge symmetry imposes constraints on the phase space for particle
motion (see section \ref{sec:2T-physics-approach} and appendix \ref{A1}).
Physical states that are invariant under this gauge symmetry reside within the
subset of phase space that satisfy these constraints. Surprisingly, all
solutions of these constraints yield nontrivial physical systems devoid of
unitarity and causality issues, provided the full phase space $(X^{M},P_{M})$,
including gauge degrees of freedom, possesses two timelike dimensions - no
less and no more.

Therefore, the appearance of an additional timelike and spacelike dimension
arises naturally from phase space gauge symmetry, rather than being
artificially imposed. Meanwhile, a plethora of gauge-invariant physical states
consistent with these constraints exists in emergent gauge-invariant effective
phase spaces in 3+1 dimensions. These are termed holographic shadows,
possessing one less timelike and one less spacelike dimension. Hence,
2T-physics in 4+2 dimensions (or more generally $d+2$) exhibits a rich
gauge-invariant physical sector mirroring the familiar one-time physics
(1T-physics) in 3+1 dimensions (or more generally $\left(  d-1\right)  +1$).

The holographic conformal shadow is just one among many shadows that are
multi-dual to one another. There is a hidden SO$(4,2)$ symmetry (and their
curved space generalizations) which prevails in all shadows. The hidden
symmetry, that is connected to partly hidden 4+2 dimensions, represents
previously unexpected non-linear symmetry properties of actions for all
shadows in 3+1 dimensions. This was discovered solely through 2T-physics
methods. What sets apart the holographic conformal shadow is its explicit
display of the Poincare symmetry linearly in flat 3+1 dimensions, along with
the SO(4,2) conformal symmetry of massless systems in a recognizable
non-linear form.

2T-physics remains consistent with all experimentally tested aspects of
1T-physics. However, it transcends 1T-physics by predicting hidden spacetime
symmetries and multi-duality relationships among a myriad of 1T-systems. These
concealed properties permeate all facets of 1T-physics, yet 1T-physics alone
lacks the capability to systematically predict them. This is where 2T-physics
furnishes new insights into the underlying 4+2-dimensional system, testable
with both 1T theory and experiments. Indeed, some of the simpler predictions
have been theoretically verified in the early stages of 2T-physics
\cite{IB+Araya}-\cite{IBQuelin}.

\subsection{Plan of this paper}

The previous research outlined above provides the foundational context for the
development of Chern-Simons (CS) terms for gravity, QCD, and QED, which are
both locally scale-invariant and geodesically complete at singularities. This
is crucial to integrate the physical implications of these CS terms into the
locally scale-invariant theory in 3+1 dimensions. Furthermore, it is
reasonable to anticipate that such CS terms in 3+1 dimensions correspond to
the holographic conformal shadow of analogous CS-type terms in 4+2-dimensional 2T-physics.

In this paper, we leverage the robust methodologies offered by 2T-physics to
investigate Chern-Simons theory. We employ 2T methods to elevate Chern-Simons
theories into frameworks that exhibit local conformal scale invariance. The
primary objectives of this endeavor are: (1) to ascertain the necessary
modifications for Chern-Simons theory to achieve local Weyl symmetry in 3+1
dimensions, and (2) to derive the Weyl symmetric 3+1 actions from the
higher-dimensional perspective of 2T-physics in 4+2 dimensions.

To address these questions, we commence with the conformally improved Standard
Model coupled to General Relativity i(SM+GR)$_{3+1}$ \cite{BST} as derived
from 2T versions of these theories in 4+2 dimensions \cite{2Tsm}%
-\cite{2TgravIB}. A pivotal prediction of 2T-physics is that the SM+GR, when
reconciled with the constraints of 2T-physics, exhibits a unique form of
hidden local conformal scale (Weyl) symmetry. Geodesic completeness follows
from this conformally invariant structure.

Given that the entire action for i(SM+GR)$_{3+1}$ is inherently locally scale
invariant (see section \ref{sec:building-weyl-symmetric}), our objective is to
determine the structural forms that the Chern-Simons terms must adopt in 3+1
dimensions to uphold the local scale symmetry of the complete i(SM+GR)$_{3+1}%
${} model. We extend the conformally improved i(SM+GR)$_{3+1}$ with
Chern-Simons terms for gravity, QCD, and QED that remain explicitly invariant
under local Weyl transformations in 3+1 dimensions. We demonstrate that
maintaining the local scale symmetry of the i(SM+GR)$_{3+1}$ model
necessitates the Pontryagin density $\tilde{R}R$ in our modified 3+1
gravitational Chern-Simons term in (\ref{RR*}) to linearly couple to a
function $f_{GR}\left(  s_{i}/\phi\right)  $ of the ratio of spin-0 fields.
Analogous scenarios arise in QCD and QED when $R_{~\sigma\mu\nu}^{\lambda}${}
is replaced by the Yang-Mills field strength $F_{\mu\nu}^{a}$ (for QCD) and
the electromagnetic field strength $F_{\mu\nu}$ (for QED).

Subsequently, we demonstrate that our conformally improved Chern-Simons terms
in 3+1 dimensions can be derived as holographic images of Chern-Simons actions
in 4+2 dimensions for gravity, QCD, and QED. We explain that the local scale
transformation in 3+1 dimensions is a remnant of 4+2 general coordinate
transformations that intertwine the extra 1+1 large dimensions with the
evident 3+1 dimensions, treated at the same footing in 2T-physics. With the
newfound local conformal symmetry, the enhanced, expanded, and geodesically
complete SM+GR+CS can be utilized to explore new physics beyond the
traditional paradigm near black hole and cosmological singularities in 3+1
dimensions at the classical level, while the 4+2 connection unveils deeper
aspects of spacetime.

The structure of the rest of this paper is as follows: In Section
\ref{sec:building-weyl-symmetric}, we elevate the gravitational Chern-Simons
term to a locally Weyl symmetric version directly in 3+1 dimensions. Next, in
Section \ref{sec:2T-physics-approach}, we illustrate how this Weyl-symmetric
Chern-Simons term emerges as a holographic image of a 2T Chern-Simons field
theoretical term in 4+2 dimensions. We expound on the interpretation of the
local scale (Weyl) symmetry of the (SM+GR+CS)$_{3+1}$ theory as a vestige of
general coordinate transformations that mix the extra 1+1 dimensions with the
3+1 dimensions. We also proceed with an examination of the QCD and QED
versions of Chern-Simons theory. In Section \ref{PandCP}, we delve into the
role of parity and CP violation in the Chern-Simons actions. Finally, in
Section \ref{sec:Discussion}, we conclude with final remarks and discuss
avenues for future research. In appendix \ref{A1}, we summarize the structure,
scope, and some of the methodologies of 2T-physics.

\section{Locally scale invariant SM, GR \& CS actions in 3+1}

\label{sec:building-weyl-symmetric}

In this section, we will promote the standard formalism of Chern-Simons
gravity defined in Eq. (\ref{RR*}) to a locally scale invariant formalism
directly in 3+1 dimensions. Our strategy is to begin with the i(SM+GR)$_{3+1}$
that is locally Weyl symmetric \cite{BST} before incorporating the
Chern-Simons correction.

\subsection{SM \& GR actions}

The existing 4+2 dimensional model (SM+GR)$_{4+2}$ \cite{2Tsm}-\cite{2TgravIB}
successfully yields i(SM+GR)$_{3+1}$ without the CS terms as a 3+1 dimensional
holographic conformal image. This theory is known to agree with the
conventional Standard Model in successfully fitting all known aspects of
particle physics down to $10^{-18}$ meters as measured by the LHC at CERN
\cite{CERN}. Since the entire action, including the CS terms, must preserve
the local Weyl symmetry, our task is to promote the CS term in (\ref{RR*}),
and the similar QCD and QED terms, to locally scale invariant versions.

In the absence of the Chern-Simons correction, the action for i(SM+GR)$_{3+1}$
is given by \cite{BST}
\begin{equation}
S_{\text{SM+GR}}^{3+1}=\int d^{4}x\sqrt{-g}\,\left(
\begin{array}
[c]{c}%
\mathcal{L}_{SM}+\frac{1}{12}\left(  \phi^{2}-2H^{\dagger}H\right)  R\left(
g\right)  \\
+\frac{1}{2}g^{\mu\nu}\left(  \partial_{\mu}\phi\partial_{\nu}\phi
-2\partial_{\mu}H^{\dagger}\partial_{\nu}H\right)  -V\left(  \phi,H\right)
\end{array}
\right)  .\label{SM-GR}%
\end{equation}
The first term, $\mathcal{L}_{SM}\left(  A_{\mu}^{\gamma,W,Z,g},\psi^{q,l}%
,\nu_{R},\chi,g_{\mu\nu},H,\phi\right)  ,$ encapsulates the familiar degrees
of freedom within the Standard Model, excluding the scalar field terms that
are displayed in (\ref{SM-GR}). $\mathcal{L}_{SM}$ encompasses the gauge
fields $A_{\mu}^{\gamma,W,Z,g}$ corresponding to the photon, $W^{\pm}$, $Z$,
and gluons $g$; the fermionic fields representing quarks and leptons
$\psi^{q,l}$; right handed neutrinos $\nu_{R}$; and candidates for dark matter
$\chi.$ These entities interact through SU$\left(  3\right)  \times$SU$\left(
2\right)  \times$U$\left(  1\right)  $ gauge symmetric Yukawa and gauge
couplings with the doublet $H$ and singlet $\phi$ (that could possibly couple
only to $\nu_{R}$ or $\chi$)$,$ and incorporating the difference between
SU$\left(  2\right)  \times$U$\left(  1\right)  $ covariant versus ordinary
derivatives in the Higgs fields's kinetic energy term $\left(  -2D_{\mu
}H^{\dagger}D^{\mu}H+2\partial_{\mu}H^{\dagger}\partial^{\mu}H\right)  $.
Moreover, all fields in $\mathcal{L}_{SM}$ are subject to minimal coupling
with the gravitational metric $g_{\mu\nu}$. The term $\mathcal{L}_{SM}$ is
written separately from the scalar sector displayed in the rest of the action
(\ref{SM-GR}) because fermion, gauge boson, and Yukawa terms in
i(SM+GR)$_{3+1}$ are already invariant under local Weyl rescalings when
minimally coupled to gravity. However for the gravitational sector and all
spin 0 fields, including the Higgs field, local scale symmetry requires the
special structures that are exhibited in the action (\ref{SM-GR}) and
discussed below.

The full action $S_{SM+GR}^{3+1}$ is invariant under local Weyl rescalings of
the form
\begin{equation}%
\begin{array}
[c]{l}%
g_{\mu\nu}\rightarrow\Omega^{2}g_{\mu\nu},\,\,\,\,\phi\rightarrow\Omega
^{-1}\phi,\,\,\,H\rightarrow\Omega^{-1}H,\,\,\,\\
\psi^{q,l}\rightarrow\Omega^{-3/2}\,\psi^{q,l},\,\,\,A_{\mu}^{\gamma
,W,Z,g}\rightarrow\Omega^{0}A_{\mu}^{\gamma,W,Z,g}%
\end{array}
\label{Wtransform}%
\end{equation}
for an arbitrary local parameter $\Omega\left(  x\right)  $.

The geodesic completeness and predictions of new physics beyond the Standard
Model emerge from the Weyl invariant special structures in this action as
discussed in section \ref{incompleteST}. These include the following features:

\begin{itemize}
\item Local scale invariance prohibits dimensionful parameters such as the
Higgs mass, cosmological constant $\Lambda,$ or the Newton constant $G_{N}$.
Consequently, the Einstein-Hilbert term cannot appear in the action
(\ref{SM-GR}). Instead, there is an effective, spacetime dependent
gravitational strength $G\left(  x\right)  $ that is determined by the scalar
singlet $\phi\left(  x\right)  $ and the doublet Higgs $H\left(  x\right)  $,
as in Eq.(\ref{Newton}), $\left(  16\pi G\left(  x\right)  \right)
^{-1}=\frac{1}{12}\left(  \phi^{2}\left(  x\right)  -2H^{\dagger}H\left(
x\right)  \right)  $.

\item In the curvature and kinetic energy terms of the scalars $\phi$ and $H,$
there is a relative minus sign. Hence $\phi$ seems to be a ghost-like field,
but there is no issue with unitarity because the local scale symmetry
eliminates the ghost by either gauge fixing it to a constant (the c-gauge,
$\phi\left(  x\right)  =\phi_{0}$) or by compensating for it in other Weyl
gauges. Without the ghostlike $\phi\left(  x\right)  $ there would be no way
to have an underlying local scale symmetry as well as the presence of a patch
of spacetime where the gravitational strength $G\left(  x\right)  $ is
positive as explained in \cite{BST} and in section \ref{DimParm}. This
relative minus sign structure between $\phi$ and $H$ first emerged in the
context of 2T-physics as outlined in section \ref{incompleteST}.

\item For the Standard Model, the potential $V\left(  \phi,H\right)  $ is the
most general renormalizable purely quartic expression,
\begin{equation}
V_{0}\left(  \phi,H\right)  =\frac{\lambda}{4}\left(  2H^{\dagger}H-\alpha
^{2}\phi^{2}\right)  ^{2}+\frac{\lambda^{\prime}}{4}\phi^{4}. \label{Vsm}%
\end{equation}
where $\alpha,\lambda,\lambda^{\prime}$ are dimensionless couplings. However,
due to renormalization effects and the coupling to gravity, $V\left(
\phi,H\right)  $ may admit additional contributions beyond $V_{0}$ that would
vanish when GR is decoupled from the SM and would be imperceptibly tiny at low
energies. Such modifications of $V\left(  \phi,H\right)  $ could become
substantial in strong gravity regions as discussed in \cite{BH+Higgs}. To
ensure the principle of local scale invariance (required if derived from
2T-physics) the full potential must be homogeneous of degree 4 under scale
transformations of the fields. Then the most general potential consistent with
Weyl symmetry is given by $V\left(  \phi,H\right)  =\phi^{4}v\left(
\sqrt{2H^{\dagger}H/\phi^{2}}\right)  $ where $v\left(  z\right)  $ is any
dimensionless function of $z,\ $as long as the deviations from the Standard
Model potential $V_{0}$ in (\ref{Vsm}) are imperceptibly small at low energies
as compared to the Planck energy scale.
\end{itemize}

The local scale symmetry (\ref{Wtransform}) can be used to remove one field
degree of freedom, such as locally rescaling $\phi\left(  x\right)  $ or
$\left(  \phi^{2}-2H^{\dagger}H\right)  ,$ or some other combination of
fields, to be spacetime independent constants in various patches of spacetime.
In choosing such gauges, one should be mindful that certain gauge choices are
valid only within geodesically incomplete patches of spacetime. Examples of
geodesically incomplete gauges are those that lead to the Einstein, Jordan, or
String frames when they are limited to only the gravity side of gravitational
singularities (see footnote \ref{EJsframes}).

In this paper we can replace the doublet $H$ by its SU$\left(  2\right)
\times$U$\left(  1\right)  $ unitary gauge fixed version $H^{\dagger}\left(
x\right)  =\left(  0,\;s\left(  x\right)  /\sqrt{2}\right)  ,$ so we
substitute everywhere $2H^{\dagger}H=s^{2}$ and $2\partial_{\mu}H^{\dagger
}\partial_{\nu}H=\partial_{\mu}s\partial_{\nu}s.$ Moreover, we suppress the
term $\mathcal{L}_{SM}$ because it plays no role in our discussion. Hence, our
starting point for new CS terms is the action in Eq.(\ref{SM-GR}) that
consists of two conformally coupled scalars $\left(  \phi,s\right)  $
interacting with gravity and each other, while obeying a local scale symmetry.

In addition to the Higgs $s\left(  x\right)  ,$ that is the only established
physical spin 0 elementary field, there may be additional spin 0 fields that
may play a role in the fundamental theory. The Weyl symmetry permits such
additional scalars or pseudoscalars that we denote by $s_{i},\,i=1,...,N$. We
can allow only 1 ghostlike $\phi$ since the Weyl symmetry is capable to ensure
unitarity only with 1 ghostlike degree of freedom. In addition to the Higgs
$s$ (renamed $s_{1}$)$,$ the $s_{i}$ can include scalar or pseudoscalar dark
matter candidates (e.g., the axion) or other spin 0 fields as motivated by
string theory (e.g., the dilaton or axion), supersymmetry, supergravity, or
Grand Unified Theories (GUTs). In the simplest Weyl invariant couplings, all
spin 0 fields are treated as conformally coupled scalars. However, there are
more general Weyl invariant couplings for multiple spin 0 fields when there
are additional scalars beyond $\phi$ and the Higgs $H$, as discussed in
\cite{BST}. Here we will deal with the case of conformally coupled scalars for
simplicity by replacing the Higgs $s$ in Eq.(\ref{SM-GR}) with $s_{i}$ and
summing over $i$ in the the kinetic and $R$ terms in (\ref{SM-GR}). Thus, we
will at times use the more general notation $\left(  \phi,s_{i}\right)  $ to
include all possible conformally coupled spin 0 fields $s_{i}$.

Likewise, the potential $V\left(  \phi,s_{i}\right)  $ can be modified to a
general function of the fields that is homogeneous of degree 4, which can be
written as $V\left(  \phi,s_{i}\right)  =\phi^{4}v\left(  s_{i}/\phi\right)  $
where $v\left(  z_{i}\right)  $ is any dimensionless function of its Weyl
invariant arguments $z_{i}\equiv s_{i}/\phi.$ Other Weyl invariant ratios,
such as $s_{i}/s_{j}$, are not independent since they can be written in terms
of $z_{i}/z_{j}$.

\subsection{CS action in GR}

We are now ready for the new Chern-Simons terms. In the gauge symmetric
version of the physically correct i(SM+GR)$_{3+1}$, we consider consistently a
Weyl invariant version of $S_{\text{$CS-GR$}}^{3+1}$. We will argue that the
Chern-Simons term $S_{\text{$CS-GR$}}^{3+1}$ in (\ref{RR*}) for gravity is
promoted to be scale invariant under Weyl transformations if the Pontryagin
density $\tilde{R}R$ linearly couples to a dimensionless function of the
ratios of the scalar fields, $f_{GR}\left(  s_{i}/\phi\right)  $ as follows
\begin{equation}
S_{\text{$CS-GR$}}^{3+1}=\int d^{4}x\,\tilde{R}R~f_{GR}\left(  s_{i}%
/\phi\right)  .\label{RRweyl}%
\end{equation}
Below we will first argue that if $f_{GR}\left(  s_{i}/\phi\right)  $ is a
function of only the ratios $s_{i}/\phi,$ or a constant, then
$S_{\text{$CS-GR$}}^{3+1}$ is invariant under \textit{global} scale
transformations. We will then argue that this is also sufficient for
$S_{\text{$CS-GR$}}^{3+1}$ to be invariant under \textit{local} scale
transformations as well.

The same reasoning applies also in the cases of QCD and QED by simply
replacing $\tilde{R}R$ by the QCD field strengths $\tilde{F}^{a}F^{a}$ or QED
field strengths $\tilde{F}F.$ Therefore, we will refrain from repeating the
arguments for QCD and QED until the discussion in section \ref{sec:QCD}.

Under the local Weyl transformations (\ref{Wtransform}), the determinant of
the metric and the curvatures transform as follows
\begin{equation}%
\begin{array}
[c]{l}%
\begin{array}
[c]{c}%
\sqrt{-g}\rightarrow\Omega^{4}\left(  x\right)  \sqrt{-g},\;\;R\left(
g\right)  \rightarrow\Omega^{-2}\left(  x\right)  \left(  R\left(  g\right)
+\text{derivatives of }\Omega\right) \\
\multicolumn{1}{l}{R_{\mu\nu}\left(  g\right)  \rightarrow\Omega^{0}\left(
x\right)  \left(  R_{\mu\nu}\left(  g\right)  +\text{derivatives of }%
\Omega\right) }\\
\multicolumn{1}{l}{R_{\mu\nu\lambda\sigma}\left(  g\right)  \rightarrow
\Omega^{2}\left(  x\right)  \left(  R_{\mu\nu\lambda\sigma}\left(  g\right)
+\text{derivatives of }\Omega\right) }\\
\multicolumn{1}{l}{R_{\;\nu\lambda\sigma}^{\mu}\left(  g\right)
\rightarrow\Omega^{0}\left(  x\right)  \left(  R_{\;\nu\lambda\sigma}^{\mu
}\left(  g\right)  +\text{derivatives of }\Omega\right) }%
\end{array}
\\
\text{hence}:\;\left(  \tilde{R}R\right)  \rightarrow1\left(  \tilde
{R}R+\text{derivatives of }\Omega\right)  .
\end{array}
\label{rescaleW}%
\end{equation}
However, the derivative terms in $\left(  \tilde{R}R\right)  \rightarrow
\left(  \tilde{R}R+\text{derivatives of }\Omega\right)  $ do not seem to
vanish unless $\Omega\left(  x\right)  $ is independent of $x$. Hence, the
Chern-Simons action $S_{\text{$CS-GR$}}^{3+1}$ in Eq.(\ref{RRweyl}) seems to
be at best invariant under \textit{global} Weyl transformations (i.e.,
transformations where $\Omega=constant$), provided the function $f_{GR}$ is
invariant under global rescalings. This is satisfied by requiring $f_{GR}$ to
be any dimensionless general function of only Weyl invariant ratios
$z_{i}=s_{i}/\phi$ as indicated in (\ref{RRweyl}).

We will now consider whether $S_{\text{$CS-GR$}}^{3+1}$ can be made invariant
under \textit{local} Weyl transformations for any local $\Omega\left(
x\right)  $. This is necessary because in the absence of local symmetry, the
ghost-like field $\phi\left(  x\right)  $ would remain as a dynamical degree
of freedom that would ruin the unitarity of the theory since it would not be
possible to remove the ghost by mapping it to a constant $\phi\left(
x\right)  \rightarrow\phi_{0}$ via a local gauge transformation. Therefore, it
is essential to determine whether $S_{\text{$CS-GR$}}^{3+1}$ can be made
invariant beyond global scale transformations, or whether it is necessary to
introduce additional terms to improve its symmetry properties.

For this purpose, we consider the Weyl tensor $C_{\mu_{1}\mu_{2}\mu_{3}\mu
_{4}}$ defined by
\begin{equation}
C_{\mu_{1}\mu_{2}\mu_{3}\mu_{4}}=\left(
\begin{array}
[c]{c}%
R_{\mu_{1}\mu_{2}\mu_{3}\mu_{4}}+\frac{1}{6}\left(  g_{\mu_{1}\mu_{3}}%
g_{\mu_{2}\mu_{4}}-g_{\mu_{1}\mu_{4}}g_{\mu_{2}\mu_{3}}\right)  ~R\\
-\frac{1}{2}\left(  R_{\mu_{1}\mu_{3}}g_{\mu_{2}\mu_{4}}-R_{\mu_{1}\mu_{4}%
}g_{\mu_{2}\mu_{3}}+R_{\mu_{2}\mu_{4}}g_{\mu_{1}\mu_{3}}-R_{\mu_{2}\mu_{3}%
}g_{\mu_{1}\mu_{4}}\right)
\end{array}
\right)  \label{Ctensor}%
\end{equation}
where $R_{\mu\nu}$ is the Ricci tensor and $R$ is the curvature scalar. The
Weyl tensor has the following properties under permutation of indices and
tracing:%
\begin{equation}%
\begin{array}
[c]{l}%
\left(  a\right)  \;C_{\mu_{1}\mu_{2}\mu_{3}\mu_{4}}=C_{\mu_{3}\mu_{4}\mu
_{1}\mu_{2}}\\
\left(  b\right)  \;C_{\mu_{1}\mu_{2}\mu_{3}\mu_{4}}=-C_{\mu_{2}\mu_{1}\mu
_{3}\mu_{4}}=-C_{\mu_{1}\mu_{2}\mu_{4}\mu_{3}}\\
\left(  c\right)  \;C_{\mu_{1}\mu_{2}\mu_{3}\mu_{4}}+C_{\mu_{1}\mu_{3}\mu
_{4}\mu_{2}}+C_{\mu_{1}\mu_{4}\mu_{2}\mu_{3}}=0,\\
\;\;\;\;\;\text{hence for any }\mu\text{ and }\nu\text{ we have: }C_{\mu
\mu_{2}\mu_{3}\mu_{4}}\varepsilon^{\nu\mu_{2}\mu_{3}\mu_{4}}=0.\\
\left(  d\right)  \text{ Traceless for any pair of indices,}\\
\;\;\;\;\;\;\text{example: }C_{\mu_{1}\mu_{2}\mu_{3}\mu_{4}}g^{\mu_{2}\mu_{3}%
}=0.
\end{array}
\label{permute}%
\end{equation}
The Riemann tensor $R_{\mu_{1}\mu_{2}\mu_{3}\mu_{4}}$ shares the properties
$\left(  a,b,c\right)  $, but the last one $\left(  d\right)  $ is valid only
for $C_{\mu_{1}\mu_{2}\mu_{3}\mu_{4}}.$

Under the local Weyl rescaling of the metric given in (\ref{rescaleW}),
$R_{\mu_{1}\mu_{2}\mu_{3}\mu_{4}},R_{\mu\nu}$, and $R$ not only rescale by an
overall factor but also have derivative terms of $\Omega\left(  x\right)  $.
The derivative terms seem to prevent $S_{\text{$CS-GR$}}^{3+1}$ from being
locally Weyl invariant. By contrast, all the derivative terms cancel in the
combination $C_{\mu_{1}\mu_{2}\mu_{3}\mu_{4}}$ given in (\ref{Ctensor}).
Therefore, the Weyl tensor only rescales by an overall factor under a
\textit{local} Weyl transformation%
\begin{equation}
C_{\mu_{1}\mu_{2}\mu_{3}\mu_{4}}\rightarrow\Omega^{2}\left(  x\right)
C_{\mu_{1}\mu_{2}\mu_{3}\mu_{4}}.
\end{equation}

Moreover, when one of the indices is raised $C_{\;\;\mu_{2}\mu_{3}\mu_{4}%
}^{\mu_{1}}=g^{\mu_{1}\nu}C_{\nu\mu_{2}\mu_{3}\mu_{4}}$, it remains fully
invariant under \textit{local} Weyl transformations%
\begin{equation}
C_{\;\;\mu_{2}\mu_{3}\mu_{4}}^{\mu_{1}}\rightarrow\Omega^{0}\left(  x\right)
~C_{\;\;\mu_{2}\mu_{3}\mu_{4}}^{\mu_{1}}.
\end{equation}
It is now evident that a locally Weyl invariant version of the CS term in
gravity is given by replacing $R_{\;\sigma\mu\nu}^{\lambda}$ with
$C_{\;\sigma\mu\nu}^{\lambda}$ in Eq.(\ref{RRweyl}),
\begin{equation}
S_{\text{$CS-GR$}}^{3+1}=\int d^{4}x\,\tilde{C}C\,f_{GR}\left(  s_{i}%
/\phi\right)  ,\,\,\,\,\,\,\tilde{C}C=\frac{1}{2}\epsilon^{\mu_{1}\mu_{2}%
\mu_{3}\mu_{4}}C_{\;\sigma\mu_{1}\mu_{2}}^{\lambda}C_{\;\lambda\mu_{3}\mu_{4}%
}^{\sigma} \label{eq:CS-CC}%
\end{equation}
where $f_{GR}\left(  s_{i}/\phi\right)  $ is any general function of the ratio
of the fields, including a constant.

It seems that $\tilde{C}C$ contains a few additional terms as compared to
$\tilde{R}R,$ thus providing the desired \textit{local} Weyl invariance
property of the action. However, using the permutation and trace properties of
the Weyl tensor listed in (\ref{permute}), it can be shown that
\begin{equation}
\tilde{C}C=\,\tilde{R}R.
\end{equation}
Hence, the local Weyl invariance of $\tilde{R}R$ can be made manifest by
rewriting it in terms of only the Weyl tensor. Therefore, the action as
written originally in Eq.(\ref{RRweyl}) is actually Weyl invariant under
\textit{local} transformations. However, this is possible only if
$f_{GR}\left(  s_{i}/\phi\right)  $ is any dimensionless function of the ratio
of the conformal scalar or pseudoscalar fields as indicated.

In conclusion, the full action for the conformally invariant (SM+GR)$_{4+2}$
augmented with the gravitational Chern-Simons correction is given by
\begin{equation}
S_{\text{SM+GR+CS}}^{3+1}=S_{\text{SM+GR}}^{3+1}+S_{\text{$CS-GR$}}%
^{3+1}.\label{ActioA2}%
\end{equation}

A similar CS term $S_{CS-QCD}^{3+1}$ associated with QCD and the axion can be
considered assuming the axion exists (see sections \ref{sec:QCD} and
\ref{PandCP}). Also in QED, the decay of the pion into two photons, and other
similar processes involving hadrons, are captured by similar terms in an
effective Lagrangian approach.

\subsection{Emergence of dimensionful parameters \label{DimParm}}

One of the virtues of this formalism is its unified approach to explain mass
generation in the Standard Model. In our theory, all dimensionful parameters
of the Standard Model and gravity are initially absent, including the Newton
constant, Higgs mass, cosmological constant, and quark, lepton, and gauge
boson masses. However - just as in spontaneously broken gauge symmetries - all
of the dimensionful parameters in the SM+GR+CS are generated from the same
single source: namely the field $\phi$ after gauge fixing the local Weyl
symmetry. This demonstrates a higher degree of unification as to the source of
all dimensionful parameters.

The gauge in which we essentially recover the usual renormalizable field
theory for the Standard Model in flat space is known as the c-gauge
\cite{BST}. In the c-gauge, the source of mass arises when the field $\phi$ is
fixed to a constant $\phi\left(  x\right)  \rightarrow\phi_{0}=\sqrt{12/(16\pi
G)}$ such that $\phi_{0}$ is associated with the emergence of the Planck
scale, $\phi_{0}\sim10^{19}$ GeV. Explicitly, for the gravitational constant
$G_{N}$, the cosmological constant $\Lambda$, and the Higgs vacuum expectation
value (VEV) $\langle\left\vert H\right\vert \rangle=v/\sqrt{2}$ that minimizes
the potential $V_{0},$ are obtained by replacing $\phi\left(  x\right)  $ by
the constant $\phi_{0}$ in the action (\ref{SM-GR})
\begin{equation}
\frac{1}{16\pi G_{N}}=\frac{\phi_{0}^{2}}{12},\,\,\,\,\,\,\frac{\Lambda}{16\pi
G_{N}}=\frac{\lambda^{\prime}}{4}\phi_{0}^{4},\,\,\,\,\,\,\,v^{2}=\alpha
^{2}\phi_{0}^{2}. \label{GLv}%
\end{equation}
Then, the electroweak symmetry breaking that is mediated by the VEV of the
Higgs doublet becomes also driven by the constant value $\phi_{0}$ of the
gauge fixed $\phi\left(  x\right)  $, so the masses of all quarks, leptons,
and gauge bosons at low energies are also driven by the same unique $\phi_{0}$ source.

In the c-gauge, we will use the letter `c' to label all fields such as,
$s_{ic}$, $g_{c}^{\mu\nu}$ etc. and $\phi_{c}\left(  x\right)  \equiv\phi
_{0},$ to emphasize that these gauge fixed fields are different than the
fields in other gauges, such as the E-gauge, the string gauge, etc. In the
c-gauge, the action given in Eq.(\ref{ActioA2}) reduces identically to the
usual Standard Model except for the curvature term that deviates from the
Einstein-Hilbert action and takes the form $\frac{1}{12}(\phi_{0}^{2}%
-s_{c}^{2}\left(  x\right)  )R\left(  g_{c}\right)  .$ This contains the
spacetime dependent effective gravitational strength $\left(  16\pi
G_{c}\left(  x\right)  \right)  ^{-1}=\frac{1}{12}(\phi_{0}^{2}-s_{c}%
^{2}\left(  x\right)  ).$ At experiments conducted at accelerators, since
energies are relatively tiny as compared to Planck energy, the dimensionless
ratio $s_{c}^{2}\left(  x\right)  /\phi_{0}^{2}$ is of order $10^{-34}.$
Accordingly, $G_{c}\left(  x\right)  $ is approximated by the constant $G_{N}$
in (\ref{GLv}) with great accuracy. Then, low energy physical phenomena are
not sensitive to the deviation from the traditional SM+GR that has a spacetime
independent gravitational constants $G_{N}.$ We see that, in the c-gauge, the
familiar Standard Model as tested and measured at low energies emerges from
Eq.(\ref{ActioA2}) almost identically.

One can compute in any other gauge. Recalling that Weyl gauge invariant
quantities are physical, agreement with the usual Standard Model conventions
and interpretations persists in regimes when the gauge invariant ratio
$s^{2}\left(  x\right)  /\phi^{2}\left(  x\right)  =s_{c}^{2}\left(  x\right)
/\phi_{0}^{2}\ll1$ is tiny.

While at low energies the locally scale invariant theory i(SM+GR)$_{3+1}$ is
practically identical to the usual SM and GR, its physical aspects are quite
different at the neighborhood of singularities such as the Big Bang and black
holes. In such spacetime regions, the Weyl symmetry repairs the geodesical
incompleteness of spacetime and introduces new physics and new perspectives of
spacetime beyond the usual Standard Model and General Relativity
\cite{BST,Cosmo2,BHoles,BH+Higgs}. The emergence of novel features becomes
conspicuous in the vicinity, at, and beyond singularities. The new aspects
apply also with supersymmetry \cite{BST}\cite{sugra4+2} and in a formalism
that repairs geodesic incompleteness in string theory \cite{BSTstrings}. Our
work in this paper extends geodesic completeness to be valid in the presence
of the Chern-Simons correction to GR as well.

\section{2T-Physics approach}

\label{sec:2T-physics-approach}

The goal in this section is to construct a 4+2 dimensional version of
Chern-Simons gravity $S_{\text{$CS-GR$}}^{4+2}$ (Eq. \ref{RRweyl}) which
reduces to the 3+1 dimensional formulation of Chern-Simons gravity
$S_{\text{$CS-GR$}}^{3+1}$ that has a local conformal scale invariance. Since
this process involves tools and techniques from 2T-physics, we include an
Appendix \ref{A1} that gives brief reviews of the main concepts, features and
tools of 2T-physics. The interested reader may consult the Appendix to better
understand the contents of this section.

In this section, we will recall the 2T-physics formulation of (SM+GR)$_{4+2}$,
construct the 4+2 dimensional Chern-Simons term $S_{\text{$CS-GR$}}^{4+2},$
and illustrate how, by gauge fixing and solving constraints, to obtain the 3+1
dimensional holographic shadow, $S_{SM+GR+CS}^{4+2}\rightarrow S_{SM+GR+CS}%
^{3+1}$ that was discussed in the previous sections.

\subsection{SM, GR \& CS actions in 4+2}

We will use the same gauge choice (i.e., the conformal shadow gauge) that was
used to derive the 2T version of the Standard Model coupled to General
Relativity in 3+1 dimensions from its 4+2 dimensional counterpart,
$S_{SM+GR}^{4+2}\rightarrow S_{SM+GR}^{3+1}$. The computations for the
reduction $S_{SM+GR}^{4+2}\rightarrow S_{SM+GR}^{3+1}$ for a $d$+2 dimensional
metric were given in \cite{2Tsm,2TgravIB,2Tgrav}. Here, we will apply similar
techniques to the reduction $S_{CS-GR}^{4+2}\rightarrow S_{CS-GR}^{3+1}$,
specializing $d+2$ to $4+2$ dimensions.

We start with the 4+2 action for 2T gravity coupled to Klein-Gordon
$\Phi,S_{i}$, Dirac $\Psi_{\alpha}^{L,R}$, and Yang-Mills $A_{M}^{a}$ type
matter fields in the absence of the Chern-Simons correction \cite{2TgravIB},
\begin{equation}%
\begin{array}
[c]{l}%
S_{SM+GR}^{4+2}=\gamma\int d^{4+2}X\,\left\{  \delta(W)~\mathcal{L}_{1}%
(\Phi,S_{i},\Psi_{\alpha}^{q,l},A^{a})\,\,+\,\,\delta^{\prime}(W)~\mathcal{L}%
_{2}(W,\Phi,S_{i})\right\}  ,\\
\mathcal{L}_{1}\equiv\sqrt{G}\left[  \frac{1}{12}\left(  \Phi^{2}-S_{i}%
^{2}\right)  R(G)+\frac{1}{2}\left(  \nabla\Phi\right)  _{G}^{2}-\frac{1}%
{2}\left(  \nabla S_{i}\right)  _{G}^{2}-V(\Phi,S_{i})+\cdots\right]  ,\\
\mathcal{L}_{2}\equiv\frac{1}{12}\sqrt{G}\left[  \left(  \Phi^{2}-S_{i}%
^{2}\right)  (4-\nabla^{2}W)+G^{MN}\nabla_{M}W~\nabla_{N}\left(  \Phi
^{2}-S_{i}^{2}\right)  \right]  .
\end{array}
\label{eq:fullS}%
\end{equation}
Here $X^{M}$ are the 4+2 spacetime coordinates, $G_{MN}\left(  X\right)  $ is
the 4+2 gravitational metric, and $R(G)$ is its Riemann curvature scalar in
4+2 from which the 3+1 dimensional shadow metric $g_{\mu\nu}\left(  x\right)
$ and its curvature scalar $R\left(  g\right)  $ are derived. Similarly,
$(\Phi,S_{i})$ with their potential $V(\Phi,S_{i}),$ are the 4+2 dimensional
fields whose 3+1 dimensional shadows $\left(  \phi\left(  x\right)
,s_{i}\left(  x\right)  \right)  $ and $V(\phi,s_{i})$ are derived. $\gamma$
is an overall normalization constant proportional to the inverse of Planck's
constant $\hbar$. The \textquotedblleft$\cdots$\textquotedblright\ in
$\mathcal{L}_{1}$ indicates additional fermions as SO(4,2)=SU(2,2) spinors
$\Psi_{\alpha}^{q,l}\left(  X\right)  $ and Yang-Mills gauge fields $A_{M}%
^{a}\left(  X\right)  $ as SO(4,2) vectors from which the 3+1 dimensional
shadow fermions $\psi^{L,R}\left(  x\right)  $ and shadow gauge fields
$A_{\mu}^{a}\left(  x\right)  $ in the i(SM+GR)$_{3+1}$ are derived.

Finally, $W\left(  X\right)  $ is an auxiliary scalar field that appears in
$\mathcal{L}_{W,\Phi,S}$ and in the delta function $\delta(W)$ and its
derivative $\delta^{\prime}(W)$ in Eq.(\ref{eq:fullS}). The origins of this
$W$ field as one of the generators of Sp$\left(  2,R\right)  $ in phase space
is explained in Appendix \ref{A2}. The restriction to the Sp$\left(
2,R\right)  $ gauge invariant sector by the vanishing of all 3 generators of
Sp$\left(  2,R\right)  $ is partially implemented by the delta function and
its derivatives. The vanishing of the other two Sp$\left(  2,R\right)  $
generators in the local field theory context, including interactions, is
described in Appendix \ref{A2}.

We now introduce the 4+2 dimensional Chern-Simons term that can be added to
the action (\ref{eq:fullS}) as an effective term induced by anomalous quantum
effects. This is the starting term in 4+2 dimensions from which $S_{CS-GR}%
^{3+1}$ in Eq.(\ref{RRweyl}) emerges as a 3+1 dimensional \textquotedblleft
shadow.\textquotedblright\ To accomplish this goal, it must have the following
form%
\begin{equation}%
\begin{array}
[c]{l}%
S_{CS-GR}^{4+2}=\gamma\int d^{4+2}X~\delta\left(  W\left(  X\right)  \right)
~\left(  \tilde{R}RA\right) \\
\left(  \tilde{R}RA\right)  \equiv\frac{1}{2}\epsilon^{M_{1}M_{2}M_{3}%
M_{4}M_{5}M_{6}}R_{\;NM_{1}M_{2}}^{M}R_{\;MM_{3}M_{4}}^{N}A_{M_{5}M_{6}}^{GR},
\end{array}
\label{action2T}%
\end{equation}
where $\epsilon^{M_{1}M_{2}M_{3}M_{4}M_{5}M_{6}}$ is the 4+2 Levi-Civita
tensor, and $R_{\;NM_{1}M_{2}}^{M}\left(  X\right)  $ is the Riemann tensor
derived from the metric $G_{MN}\left(  X\right)  $ in 4+2 dimensions. The
$A_{M_{5}M_{6}}^{GR}\left(  X\right)  $ is an antisymmetric tensor constructed
from the other 4+2 dimensional field degrees of freedom in the 2T theory.

One would like to choose some $A_{MN}^{GR}$ whose shadow would reproduce the
$f_{GR}\left(  s_{i}/\phi\right)  $ in $S_{CS-GR}^{3+1}.$ Moreover, the 3+1
dimensional curvature tensor $R_{\;\nu\mu_{1}\mu_{2}}^{\mu}\left(  g\left(
x\right)  \right)  $ constructed from the shadow metric $g_{\mu\nu}\left(
x\right)  $ in 3+1 dimensions must emerge as the shadow of its 4+2 dimensional
parent $R_{\;NM_{1}M_{2}}^{M}\left(  G\left(  X\right)  \right)  $ in 4+2
dimensions. We introduce the following $A_{MN}^{GR}$ constructed from the
derivatives of $W$ and $\Phi$, where the field $W\left(  X\right)  $ is the
one that appears in the delta function in the 2T action
\begin{equation}
A_{MN}^{GR}=\frac{1}{4}\left(  \frac{\partial W}{\partial X^{M}}\frac
{\partial\left(  \ln\Phi\right)  }{\partial X^{N}}-\frac{\partial W}{\partial
X^{N}}\frac{\partial\left(  \ln\Phi\right)  }{\partial X^{M}}\right)
f_{GR}\left(  \frac{S_{i}}{\Phi}\right)  . \label{eq:AMN-Form}%
\end{equation}
The form in Eq. (\ref{eq:AMN-Form}) is motivated on the basis of engineering
dimensions as follows. The 2T gauge symmetry \cite{2Tsm}-\cite{2Tgrav} of the
4+2 dimensional action requires the following engineering dimensions for the
coordinates $X^{M}$, the fields $W$, $\Phi$, and $S$, the metric $G_{MN}$, the
Yang-Mills gauge field $A_{M}$, the Christoffel connection $\Gamma
_{\beta\gamma}^{\alpha}$, and the Riemann tensor $R_{NM_{1}M_{2}}^{M}$:
\begin{equation}%
\begin{array}
[c]{l}%
\text{dim}(X^{M})=-1,\,\,\,\,\text{dim}(W)=-2,\,\,\,\,\,\,\,\,\,\text{dim}%
(\Phi)=+1,\,\,\,\,\,\,\,\text{dim}(S_{i})=+1,\\
\text{dim}(G_{MN})=0,\,\,\,\,\,\,\text{dim}(A_{M})=+1,\,\,\,\,\,\text{dim}%
(\Gamma_{\beta\gamma}^{\alpha})=+1,\,\,\,\,\,\text{dim}(R_{NM_{1}M_{2}}%
^{M})=+2.
\end{array}
\end{equation}
Given that the action must have dimension 0, $A_{M_{5}M_{6}}^{GR}$ should have
engineering dimension 0, after taking into account that in the volume element,
$d^{4+2}X\,\delta(W(X)),$ the factor $d^{4+2}X\,$ has dimension $-6$ while the
delta function $\delta(W(X)$ has dimension $+2.$ Accordingly, the dim$\left(
A_{MN}^{GR}\right)  =0$ is consistent with the assigned dimensions since
$\,$dim$(\partial W/\partial X^{M})=-1$ and dim$(\partial\left(  \ln
\Phi\right)  /\partial X^{M})=+1.$

\subsection{Holographic conformal shadow \label{HolShad}}

We will now outline the process by which the 4+2 dimensional metric, fields,
Riemann tensors, and antisymmetric tensor ($G_{MN},\Phi(X),S_{i}%
(X),R_{NM_{1}M_{2}}^{M},A_{MN}^{GR}$) are reduced to their 3+1 dimensional
shadow counterparts ($g_{\mu\nu}(x),\phi(x),s_{i}(x),R_{\sigma\mu_{1}\mu_{2}%
}^{\lambda}\left(  g\left(  x\right)  \right)  ,f_{GR}(s_{i}/\phi))$.

There are many ways to parameterize $X^{M}$ in 4+2 dimensions to reduce the
4+2 field theory to the emergent 3+1 Chern-Simons action. One approach is to
make a convenient choice of coordinate transformations $X^{M}\rightarrow
(w,u,x^{\mu})$ so that $W(X)=W(w,u,x^{\mu})=w$ is one of the independent
coordinates.\footnote{In flat space, we begin with $W_{\text{flat}}%
\equiv\left(  X^{2}\right)  _{\text{flat}}=X^{M}X^{N}\eta_{MN}$ where
$\eta_{MN}$ is a flat metric in 4+2 dimensions. We can always parameterize the
6 coordinates $X^{M}$ in terms of another set of 6 coordinates $\left(
w,u,x^{\mu}\right)  $ such that $X^{M}X^{N}\eta_{MN}=w.$ In curved space,
$W\left(  X\right)  $ is a general scalar field that, by the 2T gauge
transformations \cite{2Tgrav}, can be gauge fixed to $W\left(  X\right)  =w.$}
The emergent 3+1 shadows $\left(  g_{\mu\nu}(x),\phi(x),s(x)\right)  $ will be
functions of only $x^{\mu}$ and tensor indices run only in 3+1 directions,
$\mu,\nu=0,1,2,3$. The tangent basis for this curved space ($w,u,x^{\mu}$) is
given by $\partial_{M}=(\partial_{w},\partial_{u},\partial_{\mu})$ where
($w,u$) are the coordinates associated with the extra 1+1 dimensions beyond
the 3+1 coordinates $x^{\mu}$. The volume element becomes
\begin{equation}
d^{4+2}X\,~\delta(W(X))=dw\,\,du\,\,d^{4}x\,~\delta(w), \label{volume}%
\end{equation}

Our strategy is to use the fact that the 2T action is manifestly invariant
under general coordinate transformations in 4+2 dimensions to gauge fix
components of the metric as functions of ($w,u,x^{\mu}$) as shown below. In
flat spacetime, before gauge fixing, the 2T action in 4+2 has a global SO(4,2)
symmetry which is linearly realized on 4+2 coordinates $X^{M}$ and is manifest
in the action. This is the Lorentz symmetry in 4+2 dimensions which treats all
coordinates on an equal footing. In the field theoretic formulation in the
presence of gravity, the SO(4, 2) symmetry is elevated to general coordinate
invariance and Yang-Mills type gauge symmetries in 4+2 dimensions. After gauge
fixing components of the metric $G_{MN}$, there remains a general coordinate
symmetry that allows us to arbitrarily reparametrize the subspace ($u,x^{\mu
})$ without altering the gauge fixed metric $G_{MN}$. With this freedom, we
can choose a parametrization which makes use of the fact that $w$ is an
independent coordinate to ensure that the conformal shadow will only depend on
the shadow $g_{\mu\nu}\left(  x\right)  $ degrees of freedom in the end.

As outlined in Appendix \ref{A2}, the equations of motion derived from the 2T
action imposes certain kinematic constraints (called $B$ and $C$) and a
dynamical constraint (called $A$), for the metric and scalar fields $\left(
G_{MN},\Phi,S_{i}\right)  $. The kinematic $B\&C$ constraints, combined with
gauge conditions for the metric $G_{MN}\left(  X\right)  $ and its inverse
$G^{MN}\left(  X\right)  $ are solved by the following tensor configuration as
functions of $\left(  w,u,x^{\mu}\right)  $ \cite{2Tgrav}
\begin{equation}
G_{MN}\left(  X\right)  =\overset{%
\begin{array}
[c]{llll}%
\;M/N\;\; & \;w\; & \;u\;\; & \;\;\nu\;\;\;\;\;\;\;\;\;\;
\end{array}
}{%
\begin{array}
[c]{l}%
w\\
u\\
\mu
\end{array}
\left(
\begin{array}
[c]{ccc}%
0 & -1 & 0\\
-1 & -4w & 0\\
0 & 0 & G_{\mu\nu}\left(  X\right)
\end{array}
\right)  ,}\text{ }G^{MN}\left(  X\right)  =\overset{%
\begin{array}
[c]{llll}%
\;M/N\;\; & \;w\; & \;u\;\; & \;\;\nu\;\;\;\;\;\;\;\;\;\;
\end{array}
}{%
\begin{array}
[c]{l}%
w\\
u\\
\mu
\end{array}
\left(
\begin{array}
[c]{ccc}%
4w & -1 & 0\\
-1 & 0 & 0\\
0 & 0 & G^{\mu\nu}\left(  X\right)
\end{array}
\right)  ,} \label{42metric}%
\end{equation}
where
\begin{equation}%
\begin{array}
[c]{l}%
G_{\mu\nu}\left(  X\right)  =e^{-4u}\tilde{g}_{\mu\nu}\left(  x,we^{4u}%
\right)  ,\\
\Phi(X)=e^{2u}\,\tilde{\phi}(x,we^{4u}),\;\;S_{i}(X)=e^{2u}\,\tilde{s}%
_{i}(x,we^{4u}).
\end{array}
\label{solutionConstr}%
\end{equation}
Because of the delta function in the volume element in Eq. (\ref{volume}), we
can consider a Kaluza-Klein type series expansion in powers of $w$,
\begin{equation}%
\begin{array}
[c]{l}%
G_{\mu\nu}\left(  X\right)  =e^{-4u}\left(  g_{\mu\nu}\left(  x\right)
+we^{4u}g_{1\mu\nu}\left(  x\right)  +\frac{1}{2}(we^{4u})^{2}g_{2\mu\nu
}\left(  x\right)  +\cdots\right) \\
\Phi(X)=e^{2u}\left(  \phi(x)+we^{4u}\phi_{1}(x)+\frac{1}{2}(we^{4u})^{2}%
\phi_{2}(x)+\cdots\right) \\
S_{i}(X)=e^{2u}\left(  s_{i}(x)+we^{4u}s_{1i}(x)+\frac{1}{2}(we^{4u}%
)^{2}s_{2i}(x)+\cdots\right)  .
\end{array}
\label{expand}%
\end{equation}
The lowest modes in this expansion are the 3+1 dimensional shadows of the 4+2
dimensional fields, $\left(  G_{MN}(X),\Phi(X),S_{i}(X)\right)  \rightarrow
\left(  g_{\mu\nu}\left(  x\right)  ,\phi(x),s_{i}\left(  x\right)  \right)
.$ The additional fields in the $w$-expansion are called prolongations of the
shadow for each field.

It was established in \cite{2Tgrav} that the prolongations are not independent
but are all determined by the shadows $\left(  g_{\mu\nu}\left(  x\right)
,\phi(x),s_{i}\left(  x\right)  \right)  $ up to gauge freedom. The dynamics
of the shadows $\left(  g_{\mu\nu}\left(  x\right)  ,\phi(x),s_{i}\left(
x\right)  \right)  $ (given by the $A$ equations in Appendix \ref{A2}) as
derived from the parent 2T action $(SM+GR+CS)_{4+2}$ are also reproduced by
the 3+1 dimensional shadow action $(SM+GR+CS)_{3+1}$ . The holographic
conformal shadow action in 3+1 looks like the familiar relativistic field
theory with one additional requirement: namely that it is conformally scale
invariant in the special form exhibited in (\ref{SM-GR}) and (\ref{RRweyl}).

We now turn to the similar derivation of the conformal shadow for the case of
the CS action $S_{CS-GR}^{4+2}\rightarrow S_{CS-GR}^{3+1}.$ Using the
expansions in powers of $w$ for $\left(  G_{MN}(X),\Phi(X),S_{i}(X)\right)  $
given above, we need to evaluate the derivatives of these fields in order to
compute $R_{\;NM_{1}M_{2}}^{M}$and $A_{M_{5}M_{6}}^{GR}.$ In particular, for
$A_{M_{5}M_{6}}^{GR}$ one must evaluate $\partial_{M}W$ and $\partial_{M}%
(\ln{\Phi})=$ $\partial_{M}\ln\phi(x,we^{4u})$ by using the chain rule. Since
$W(X)=W(w,u,x)=w$ is a function of only the coordinate $w$, the derivative
$\partial_{M}W$ vanishes for directions $M$ except for $M=w$
\begin{equation}
\partial_{M}W(X)=\frac{\partial w}{\partial X^{M}}\frac{\partial}{\partial
w}W(w,u,x)=\delta_{M}^{w}.
\end{equation}
Similarly,
\begin{equation}%
\begin{array}
[c]{l}%
\partial_{N}\ln({\Phi})=\left(  \frac{\partial w}{\partial X^{N}}%
\frac{\partial}{\partial w}+\frac{\partial u}{\partial X^{N}}\frac{\partial
}{\partial u}+\frac{\partial x^{\mu}}{\partial X^{N}}\frac{\partial}{\partial
x^{\mu}}\right)  \ln(e^{2u}\tilde{\phi}(x,we^{4u}))\\
\;\;\;=\left(  2\delta_{N}^{u}+\frac{\tilde{\phi}^{\prime}(x,we^{4u})}%
{\tilde{\phi}(x,we^{4u})}e^{4u}\left(  \delta_{N}^{w}+4w\delta_{N}^{u}\right)
+\delta_{N}^{\mu}\frac{\partial_{\mu}\tilde{\phi}(x,we^{4u})}{\tilde{\phi
}(x,we^{4u})}\right)  .
\end{array}
\end{equation}
where $\tilde{\phi}^{\prime}(x,we^{4u})$ is the derivative of $\tilde{\phi}$
with respect to the argument $we^{4u}.$ The resulting expression for
$A_{MN}^{GR}$ is
\[
A_{MN}^{GR}=\frac{1}{4}\left(  \delta_{M}^{w}\partial_{N}(\ln{\Phi}%
)\,-\delta_{N}^{w}\partial_{M}(\ln{\Phi})\right)  ~f_{GR}\left(  \frac{S_{i}%
}{\Phi}\right)  .
\]
Components of the asymmetric tensor $A_{MN}^{GR}$ that do not contain $M=w$ or
$N=w$ vanish because of the factor $\delta_{M}^{w}.$ Moreover, the component
$A_{ww}$ also vanishes because of the antisymmetry in $M\leftrightarrow N$
\begin{equation}
A_{ww}^{GR}=0,\;A_{uu}^{GR}=0,\;A_{u\nu}^{GR}=-A_{\mu u}^{GR}=0\text{ and
}A_{\mu\nu}^{GR}=0.
\end{equation}

The vanishing condition $W(X)=w\rightarrow0$ due to the delta function should
be implemented only after evaluating all the derivatives $\partial_{w}$ with
respect to $w,$ as done above. The remaining components of $A_{MN}^{GR}$ in
the limit of $w\rightarrow0$ are%
\begin{equation}%
\begin{array}
[c]{l}%
A_{wu}^{GR}=-(A_{uw}^{GR})=\frac{1}{4}\left(  2+4we^{4u}\frac{\tilde{\phi
}^{\prime}(x,we^{4u})}{\tilde{\phi}(x,we^{4u})}\right)  f_{GR}\left(
\frac{\tilde{s}_{i}(x,we^{4u})}{\tilde{\phi}(x,we^{4u})}\right)
\overset{w\rightarrow0}{\rightarrow}\frac{1}{2}f_{GR}\left(  \frac{s_{i}%
(x)}{\phi(x)}\right)  ,\\
A_{w\nu}^{GR}=-(A_{\mu w}^{GR})=\frac{1}{4}\frac{\partial_{\mu}\tilde{\phi
}(x,we^{4u})}{\tilde{\phi}(x,we^{4u})}f_{GR}\left(  \frac{\tilde{s}%
_{i}(x,we^{4u})}{\tilde{\phi}(x,we^{4u})}\right)  \overset{w\rightarrow
0}{\rightarrow}\frac{1}{4}\frac{\partial_{\mu}\phi(x)}{\phi(x)}f_{GR}\left(
\frac{s_{i}(x)}{\phi(x)}\right)  .
\end{array}
\end{equation}
Inserting this form of $A_{M_{5}M_{6}}^{GR}$ in the $S_{CS-GR}^{4+2}$ action
in Eq. (\ref{action2T}), we obtain%
\begin{equation}%
\begin{array}
[c]{l}%
\left(  \tilde{R}RA\right)  =\frac{1}{2}\varepsilon^{M_{1}M_{2}M_{3}M_{4}%
M_{5}M_{6}}R_{\;NM_{1}M_{2}}^{M}R_{\;MM_{3}M_{4}}^{N}A_{M_{5}M_{6}}^{GR}\\
\;\;\;=\frac{1}{2}\varepsilon^{M_{1}M_{2}M_{3}M_{4}wu}R_{\;NM_{1}M_{2}}%
^{M}R_{\;MM_{3}M_{4}}^{N}2A_{wu~}^{GR}-\frac{1}{2}\varepsilon^{M_{1}M_{2}%
M_{3}M_{4}\mu_{5}w}R_{\;NM_{1}M_{2}}^{M}R_{\;MM_{3}M_{4}}^{N}2A_{\mu_{5}%
w}^{GR}.
\end{array}
\end{equation}
Because of antisymmetry properties of the Levi-Civita tensor, in the first
term $M_{1}M_{2}M_{3}M_{4}$ must all point in the 3+1 directions $\mu_{1}%
\mu_{2}\mu_{3}\mu_{4},$ while in the second term at least one of $M_{1}%
M_{2}M_{3}M_{4}$ must point in the $u$ direction. Therefore, we can simplify
\begin{equation}
\left(  \tilde{R}RA\right)  =\frac{1}{2}\varepsilon^{\mu_{1}\mu_{2}\mu_{3}%
\mu_{4}}~\left(  R_{\;N\mu_{1}\mu_{2}}^{M}R_{\;M\mu_{3}\mu_{4}}^{N}%
~2A_{wu~}-R_{\;N\mu_{1}\mu_{2}}^{M}R_{\;M\mu_{3}u}^{N}8A_{\mu_{5}w}%
^{GR}\right)  . \label{RRA2}%
\end{equation}

The next task is to compute the components of the Riemann tensors
$R_{NM_{1}M_{2}}^{M}\left(  G\right)  ,\,R_{MM_{3}M_{4}}^{N}\left(  G\right)
$ for the 4+2 metric $G_{MN}$ given in Eq.(\ref{42metric}) and express the
result in terms of the shadow $g_{\mu\nu}\left(  x\right)  $ and the
prolongations $g_{1\mu\nu}\left(  x\right)  ,g_{2\mu\nu}\left(  x\right)  ,$
etc. as fields in 3+1 dimensions. Borrowing from Eqs.(6.10-6.11) in
\cite{2Tgrav}, we list the non-vanishing components of $R_{~NM_{1}M_{2}}%
^{M}\left(  G\right)  $ with one upper index in the limit $w\rightarrow0$
\begin{equation}%
\begin{array}
[c]{l}%
R_{\;\rho\mu_{1}w}^{u}(G)=\frac{e^{4u}}{4}\left(  g_{1\mu_{1}}^{\sigma
}g_{1\sigma\rho}-2g_{2\mu_{1}\rho}\right)  ,\\
R_{\;\rho\mu_{1}\mu_{2}}^{u}(G)=\frac{1}{2}\left(  \nabla_{\mu_{1}}g_{1\mu
_{2}\rho}-\nabla_{\mu_{2}}g_{1\mu_{1}\rho}\right)  ,\\
R_{\;w\mu_{1}w}^{\lambda}(G)=\frac{e^{8u}}{4}\left(  g_{1}^{\lambda\sigma
}g_{1\sigma\mu_{1}}-2g_{2\mu_{1}}^{\lambda}\right)  ,\\
R_{\;w\mu_{1}\mu_{2}}^{\lambda}(G)=\frac{e^{4u}}{2}\left(  \nabla_{\mu_{1}%
}g_{1\mu_{2}}^{\lambda}-\nabla_{\mu_{2}}g_{1\mu_{1}}^{\lambda}\right)  ,\\
R_{\;\rho\mu_{1}w}^{\lambda}(G)=\frac{e^{4u}}{2}g^{\lambda\sigma}\left(
\nabla_{\sigma}g_{1\rho\mu_{1}}-\nabla_{\rho}g_{1\sigma\mu_{1}}\right) \\
R_{\;\rho\mu_{1}\mu_{2}}^{\lambda}(G)=\left(  R_{\;\rho\mu_{1}\mu_{2}%
}^{\lambda}(g)-g_{1[\mu_{1}}^{\lambda}g_{\mu_{2}]\rho}-\delta_{\lbrack\mu_{1}%
}^{\lambda}g_{1\mu_{2}]\rho}\right)  .
\end{array}
\label{curvs}%
\end{equation}
These expressions include the prolongations of the shadow that survive the
$w\rightarrow0$ limit after taking $\partial_{w}$ derivatives. Here
$\nabla_{\mu}$ is the covariant derivative with respect to the 3+1 shadow
metric $g_{\mu\nu}\left(  x\right)  $ and $R_{\;\rho\mu_{1}\mu_{2}}^{\lambda
}(g)$ is its Riemann tensor, but the shadow $R_{\;\rho\mu_{1}\mu_{2}}%
^{\lambda}(g)$ is only the first part of $R_{\;\rho\mu_{1}\mu_{2}}^{\lambda
}(G)$ as seen in the last term of Eq.(\ref{curvs}). Moreover, any upper index
on the prolongations, such as $g_{1\mu_{1}}^{\lambda},$ $g_{1}^{\lambda\sigma
}$ or $g_{2\mu_{1}}^{\lambda},$ was obtained by using the inverse shadow
metric $g^{\lambda\sigma},$ such as $g_{1\mu_{1}}^{\lambda}\equiv
g^{\lambda\sigma}g_{1\sigma\mu_{1}},$ etc.

Inserting these results in Eq.(\ref{RRA2}), we see that in the sums over $M$
and $N$, only the last curvature $R_{\;\rho\mu_{1}\mu_{2}}^{\lambda}(G)$
contributes, so the second term in (\ref{RRA2}) drops out, and the first term
yields: $\left(  \tilde{R}RA\right)  =\frac{1}{2}\varepsilon^{\mu_{1}\mu
_{2}\mu_{3}\mu_{4}}R_{\;\rho\mu_{1}\mu_{2}}^{\lambda}\left(  G\right)
R_{\;\lambda\mu_{3}\mu_{4}}^{\rho}\left(  G\right)  ~f_{GR}\left(  s_{i}%
/\phi\right)  .$ In this expression, $R_{\;\rho\mu_{1}\mu_{2}}^{\lambda
}\left(  G\right)  $ given in (\ref{curvs}) includes contributions from the
prolongation $g_{1\mu\nu}$ and $g_{2\mu\nu}$ in Eq.(\ref{curvs}). However,
these pieces drop out in the sums over $\mu_{1},\mu_{2},\mu_{3},\mu_{4}$ due
to the complete antisymmetric nature of $\varepsilon^{\mu_{1}\mu_{2}\mu_{3}%
\mu_{4}}$ versus the symmetric nature of $g_{1\mu\nu}$ and $g_{2\mu\nu}.$ In
the shadow action $S_{CS-GR}^{3+1}$, only the shadow fields $\left(  g_{\mu
\nu}\left(  x\right)  ,\phi(x),s_{i}\left(  x\right)  \right)  $ survive while
all prolongations drop out. Thus, the final result contains only the shadow
$R_{\;\rho\mu_{1}\mu_{2}}^{\lambda}(g)$ piece of $R_{\;\rho\mu_{1}\mu_{2}%
}^{\lambda}(G)$
\begin{equation}
\left(  \tilde{R}RA\right)  =\frac{1}{2}\varepsilon^{\mu_{1}\mu_{2}\mu_{3}%
\mu_{4}}R_{\;\rho\mu_{1}\mu_{2}}^{\lambda}\left(  g\right)  ~R_{\;\lambda
\mu_{3}\mu_{4}}^{\rho}\left(  g\right)  ~f_{GR}\left(  s_{i}/\phi\right)  .
\end{equation}

At this point, the density $\left(  \tilde{R}RA\right)  $ is independent of
$u$ as well as $w$. Therefore, in the volume element in Eq.(\ref{volume}), the
$u$ and $w$ integrations are performed, and the resulting constant coefficient
is cancelled against $\gamma$ as done in every term of the full action
\cite{2TgravIB}, to obtain the conformal shadow action in 3+1 dimensions
\begin{equation}%
\begin{array}
[c]{l}%
S_{CS-GR}^{4+2}=\gamma\int d^{4+2}X~\delta\left(  W\left(  X\right)  \right)
~\frac{1}{2}\epsilon^{M_{1}M_{2}M_{3}M_{4}M_{5}M_{6}}R_{\;NM_{1}M_{2}}%
^{M}R_{\;MM_{3}M_{4}}^{N}A_{M_{5}M_{6}}^{GR}\\
\overset{\text{Shadow}}{\Longrightarrow}S_{CS-GR}^{3+1}=\int d^{4}x\,\frac
{1}{2}\varepsilon^{\mu_{1}\mu_{2}\mu_{3}\mu_{4}}R_{\;\rho\mu_{1}\mu_{2}%
}^{\lambda}\left(  g\right)  ~R_{\;\lambda\mu_{3}\mu_{4}}^{\rho}\left(
g\right)  ~f_{GR}\left(  s_{i}/\phi\right)  .
\end{array}
\label{Emergence}%
\end{equation}
This proves that the 3+1 Chern-Simons action term in Eq.(\ref{RRweyl}) indeed
emerges as a holographic image of the 4+2 action term in Eq.(\ref{action2T}).

It should be emphasized that the 3+1 dimensional shadow degrees of freedom are
self-sufficient to describe the gauge invariant physical phenomena in 3+1
dimensions. However, as seen in Eq.(\ref{curvs}) there are prolongations of
the shadow that describe non-trivial phenomena occurring in the extra 1+1
dimensions. The prolongations of the metric $g_{1\mu\nu}\left(  x\right)  $
and $g_{2\mu\nu}\left(  x\right)  $ are fully determined by the shadow
$g_{\mu\nu}$ as discussed in \cite{2TgravIB}. Therefore, the shadow determines
all the prolongations of the geometry shown in (\ref{curvs}) and similarly for
the prolongations of all other field degrees of freedom $\Phi,S_{i},\Psi
,A_{M}^{a}.$

The fact that there are (Kaluza-Klein type) prolongations in the extra
dimensions as seen in Eq.(\ref{curvs}) is also noted for all shadows (beyond
the conformal shadow discussed in this section). This is also a common feature
in the classical and quantum phase spaces in various shadows of a single
particle in 4+2 dimensions as displayed in \cite{Survey}\cite{IB+Araya}%
\cite{DualFieldThs}. All of this, including the multi-dualities among the
shadows, are indirect indications of the existence of the extra dimensions.
All prolongations are determined by the shadows, so unlike the usual
Kaluza-Klein setting the prolongations are not independent degrees of freedom.
However, there may be ways to analyze physical effects associated with the
prolongations, thus probing the extra dimensions directly.

\subsection{The emergent Weyl symmetry in 3+1 \label{origins}}

\label{Wsymm}

We seize this moment to elucidate the 4+2 genesis of local conformal scale
(Weyl) symmetry within the conformal shadow. Recall that the 4+2 parent
actions, denoted as $S_{CS-GR}^{4+2}$ and $S_{SM+GR}^{4+2}$, are invariant
under general coordinate transformations in 4+2 dimensions, albeit lacking
Weyl symmetry. During the process of gauge fixing and solving the kinematic
constraints, a portion of the general coordinate reparametrizations is also
fixed. Nevertheless, within the conformal shadow, alongside the residual
reparametrization symmetry, there persists an additional symmetry stemming
from reparametrizations that mix the extra 1+1 dimensions with the 3+1
dimensions. This residual 4+2 symmetry manifests as the 3+1 Weyl
transformations delineated in Eq.(\ref{rescaleW}).

To illustrate this, let us examine the solutions of the kinematic B\&C
constraints for $G_{MN}\left(  w,u,x\right)  ,\Phi\left(  w,u,x\right)  $, and
$S_{i}\left(  w,u,x\right)  $ as provided in Eqs.(\ref{solutionConstr}%
,\ref{expand}). We apply a specific general coordinate transformation that
involves mixing $u$ with $x^{\mu}$ as follows: $u\rightarrow\left(
u-\lambda\left(  x\right)  /2\right)  $ where $\lambda\left(  x\right)  $
represents an arbitrary function of the 3+1 spacetime coordinates $x^{\mu}$.
Under this transformation, the fields transform as follows: $\Phi\left(
w,u,x\right)  \rightarrow\Phi\left(  w,\left(  u-\frac{1}{2}\lambda\left(
x\right)  \right)  ,x\right)  $, and similarly for $S_{i}\left(  w,u,x\right)
$. While $G_{MN}\left(  w,u,x\right)  $ typically transforms as a tensor,
under this particular transformation, the gauge-fixed form in (\ref{42metric})
remains unchanged. Instead, only $g_{\mu\nu}\left(  w,u,x\right)  $ transforms
into $g_{\mu\nu}\left(  w,\left(  u-\frac{1}{2}\lambda\left(  x\right)
\right)  ,x\right)  $. By substituting the $w$ series expansion of Eq.
(\ref{expand}), we obtain:%
\begin{equation}%
\begin{array}
[c]{l}%
\Phi\left(  w,u,x\right)  \rightarrow\Phi\left(  w,\left(  u-\frac{1}%
{2}\lambda\left(  x\right)  \right)  ,x\right) \\
\;\;=e^{2u-\lambda\left(  x\right)  }\left(  \phi(x)+we^{4u-2\lambda\left(
x\right)  }\phi_{1}(x)+\frac{1}{2}(we^{4u-2\lambda\left(  x\right)  })^{2}%
\phi_{2}(x)+\cdots\right)
\end{array}
\end{equation}
This is equivalent to transforming the shadow and prolongation fields $\left(
\phi(x),\phi_{1}(x),\phi_{2}(x),\cdots\right)  $ including all higher modes as
follows%
\begin{equation}%
\begin{array}
[c]{llll}%
\phi(x)\rightarrow\phi(x)e^{-\lambda\left(  x\right)  }, & \phi_{1}%
(x)\rightarrow\phi_{1}(x)e^{-3\lambda\left(  x\right)  }, & \phi
_{2}(x)\rightarrow\phi_{2}(x)e^{-5\lambda\left(  x\right)  }, & \cdots\\
s(x)\rightarrow s(x)e^{-\lambda\left(  x\right)  }, & s_{1}(x)\rightarrow
s_{1}(x)e^{-3\lambda\left(  x\right)  }, & s_{2}(x)\rightarrow s_{2}%
(x)e^{-5\lambda\left(  x\right)  }, & \cdots\\
g_{\mu\nu}(x)\rightarrow g_{\mu\nu}(x)e^{2\lambda\left(  x\right)  }, &
g_{1\mu\nu}(x)\rightarrow g_{1\mu\nu}e^{0~\lambda\left(  x\right)  }, &
g_{2\mu\nu}(x)\rightarrow g_{2\mu\nu}e^{-2\lambda\left(  x\right)  }, & \cdots
\end{array}
\label{leftover}%
\end{equation}
These Weyl-type local scale transformations in $3+1$ dimensions are assured to
be symmetries of the parent actions $S_{CS-GR}^{4+2}$ and $S_{SM+GR}^{4+2}$ as
they correspond to specific general coordinate transformations in 4+2
dimensions. It becomes apparent that the first column in Eq.(\ref{leftover})
represents the Weyl transformation of the shadow fields $\left(  \phi
,s,g_{\mu\nu}\right)  $, constituting the local scale symmetry of the shadow
actions $S_{CS-GR}^{3+1}$ and $S_{SM+GR}^{3+1}.$ This demonstrates that the
local Weyl symmetry in 3+1 relativistic field theory in Eq.(\ref{Wtransform})
is synonymous with a distinct general coordinate transformation in 4+2
dimensions, $\Omega\left(  x\right)  =e^{\lambda\left(  x\right)  }$,
entwining the extra 1+1 dimensions $u,w$ with the 3+1 dimensions $x^{\mu}$
while keeping $w=0.$

\subsection{4+2 dimensional QCD \& QED CS actions}

\label{sec:QCD}

In this section, we discuss the QCD and QED formalisms of Chern-Simons theory.
In \cite{2Tsm}, it was demonstrated that Weyl symmetric QCD action emerges
from the action $S_{SM+GR}^{4+2}$ given by Eq.(\ref{eq:fullS}) in the absence
of the Chern-Simons corrections. Therefore, constructing a locally scale
invariant formalism of Chern-Simons theory in QCD amounts to determining the
structural forms required for the modified QCD Chern-Simons term to preserve
the local scale invariance of the full action.

It is known from 1T-physics that the QCD formalism of Chern-Simons theory is
structurally similar to its gravitational counterpart. We will exploit this
correspondence to construct the QCD Chern-Simons term. For QCD the
pseudoscalar axion field $a\left(  x\right)  $, which is one of the possible
fields in $s_{i}\left(  x\right)  $ in the i(SM+GR)$_{3+1}$, couples to the
QCD instanton density $\tilde{F}F$ instead of the Pontryagin density
$\tilde{R}R$. By extension of the arguments presented in section
\ref{sec:building-weyl-symmetric}, we expect that a locally scale invariant
QCD Chern-Simons term in 3+1 dimensions will involve a function $f_{QCD}%
(a/\phi)$ linearly coupled to the instanton density $\tilde{F}F$:
\begin{equation}
S_{\text{$CS-QCD$}}^{3+1}=\int d^{4}x\,f_{QCD}(a/\phi)\,\tilde{F}%
F,\,\,\,\,\,\,\,\tilde{F}F=\frac{1}{2}\epsilon^{\mu_{1}\mu_{2}\mu_{3}\mu_{4}%
}F_{\;\mu_{1}\mu_{2}}^{a}F_{\;\mu_{3}\mu_{4}}^{b}\eta_{ab}.\label{SQCD31}%
\end{equation}
where $\eta_{ab}$ is the Killing metric for the gauge group SU$\left(
3\right)  $, and $F_{\;\mu_{1}\mu_{2}}^{a}$ is the QCD Yang-Mills field
strength. To match to previous work on this case \cite{PQ}, we should take a
linear function of the axion field $a$,
\begin{equation}
f_{QCD}(a/\phi)=\frac{a}{\phi}c,\label{fQCD}%
\end{equation}
where $c$ is a dimensionless constant. The field $\phi\left(  x\right)  $
turns into a constant $\phi\left(  x\right)  \rightarrow\phi_{0}$ when the
local Weyl symmetry is gauge fixed to produce the familiar term \cite{PQ} in
the previously Weyl non-invariant SM+GR. This structure shows the hidden
conformal symmetry including the Weyl symmetry and its relation to the
underlying 4+2 dimensional spacetime.

We now consider how this 3+1 dimensional QCD Chern-Simons term can be derived
from a 4+2 dimensional parent term. By extension of the arguments presented in
section \ref{sec:2T-physics-approach}, the 4+2 dimensional parent is expected
to take the form
\begin{equation}%
\begin{array}
[c]{l}%
S_{\text{$CS-QCD$}}^{4+2}=\int d^{4+2}X~\delta\left(  W\left(  X\right)
\right)  \;\left(  \tilde{F}FA\right)  ,\;\;\;\\
\left(  \tilde{F}FA\right)  \equiv\frac{1}{2}\epsilon^{M_{1}M_{2}M_{3}%
M_{4}M_{5}M_{6}}\eta_{ab}F_{\;M_{1}M_{2}}^{a}F_{\;M_{3}M_{4}}^{b}A_{M_{5}%
M_{6}}^{QCD}.
\end{array}
\label{actionQCD}%
\end{equation}
The $F_{\;M_{1}M_{2}}^{a}$ is given by
\begin{equation}
F_{M_{1}M_{2}}^{a}=\frac{\partial A_{M_{2}}^{a}}{\partial X^{M_{1}}}%
-\frac{\partial A_{M_{1}}^{a}}{\partial X^{M_{2}}}+gf_{~~bc}^{a}A_{M_{1}}%
^{b}A_{M_{2}}^{c}%
\end{equation}
where $A_{M}^{a}\left(  X\right)  $ is the 4+2 parent of the 3+1 Yang-Mills
gauge field $A_{\mu}^{a}\left(  x\right)  $. The $A_{M_{5}M_{6}}^{QCD}$ in
Eq.(\ref{actionQCD}) is given by
\begin{equation}
A_{M_{5}M_{6}}^{QCD}=\frac{1}{2}\left(  \frac{\partial W}{\partial X^{M_{5}}%
}\frac{\partial\left(  \ln\Phi\right)  }{\partial X^{M_{6}}}-\frac{\partial
W}{\partial X^{M_{6}}}\frac{\partial\left(  \ln\Phi\right)  }{\partial
X^{M_{5}}}\right)  f_{QCD}\left(  \frac{\mathcal{A}}{\Phi}\right)
\label{eq:AMN-QCD}%
\end{equation}
where $\mathcal{A}\left(  X\right)  $ is the 4+2 parent of the 3+1
pseudoscalar axion field $a\left(  x\right)  $ that appears in $S_{QCD}%
^{3+1}.$ We can follow a similar procedure as outlined in section
\ref{sec:2T-physics-approach} to reduce the 4+2 dimensional fields, field
strength tensors, and antisymmetric tensor ($\Phi,\mathcal{A},F_{M_{1},M_{2}%
}^{a},A_{MN}^{QCD}$) to their 3+1 dimensional counterparts $\left(
\phi,a,F_{\mu_{1}\mu_{2}}^{a},f_{QCD}(a/\phi)\right)  $. This procedure yields
the action in Eq.(\ref{SQCD31}) as a holographic image of Eq.(\ref{actionQCD}).

Similar considerations apply to anomalous terms in QED where instead of the
axion $a$, we have the neutral pion $\pi^{0}$ (or similar relevant hadrons)
and instead of the QCD field strength $F_{M_{1}M_{2}}^{a}\left(  X\right)  $,
we have the QED field strength $F_{M_{1}M_{2}}\left(  X\right)  .$

\section{P and CP violation}

\label{PandCP}

We will now discuss the parity (P) and charge parity (CP) conjugation
properties of our locally scale invariant Chern-Simons action. In 1T-physics,
the gravitational, QCD and QED Chern-Simons actions terms are expected to
satisfy certain transformation properties under parity and CP transformations.
The goal of this section is to determine how the functions $f_{GR}$, $f_{QCD}$
and $f_{QED}$ need to be restricted for consistency with the established 1T
transformation properties.

We previously obtained the following Weyl invariant 3+1 Chern-Simons action
terms for gravity, QCD and QED in Eqs.(\ref{RRweyl},\ref{SQCD31}):%
\begin{equation}%
\begin{array}
[c]{ll}%
S_{\text{$CS-GR$}}^{3+1}=\int d^{4}x\,f_{GR}\left(  s_{i}/\phi\right)
\,\tilde{R}R\;,\; & \tilde{R}R=\frac{1}{2}\epsilon^{\mu_{1}\mu_{2}\mu_{3}%
\mu_{4}}R_{\;\sigma\mu_{1}\mu_{2}}^{\lambda}R_{\;\lambda\mu_{3}\mu_{4}%
}^{\sigma},\\
S_{\text{$CS-QCD$}}^{3+1}=\int d^{4}x\,\,f_{QCD}(a/\phi)\,\left(  \tilde
{F}F\right)  _{QCD}, & \left(  \tilde{F}F\right)  _{QCD}=\frac{1}{2}%
\epsilon^{\mu_{1}\mu_{2}\mu_{3}\mu_{4}}F_{\mu_{1}\mu_{2}}^{a}F_{\mu_{3}\mu
_{4}}^{b}\eta_{ab},\\
S_{\text{$CS-QED$}}^{3+1}=\int d^{4}x\,\,f_{QED}(\pi^{0}/\phi)\,\left(
\tilde{F}F\right)  _{QED},\;\; & \left(  \tilde{F}F\right)  _{QED}=\frac{1}%
{2}\epsilon^{\mu_{1}\mu_{2}\mu_{3}\mu_{4}}F_{\mu_{1}\mu_{2}}F_{\mu_{3}\mu_{4}%
}.
\end{array}
\label{grQCDQED}%
\end{equation}
In 3+1 dimensions, $\tilde{R}R,(\tilde{F}F)_{QCD},(\tilde{F}F)_{QED}$ flip
signs under parity (P) and charge plus parity (CP) transformations. Hence, the
P and CP transformation characteristics of the aforementioned CS action terms
hinge crucially on the nature of the functions $f_{GR}\left(  s_{i}%
/\phi\right)  $, $f_{QCD}\left(  a/\phi\right)  ,$ and $f_{QED}\left(  \pi
^{0}/\phi\right)  $. Notably, $f_{QED}\left(  \pi^{0}/\phi\right)  $ is a
recognized P and CP-odd function, as detailed below, rendering
$S_{\text{$CS-QED$}}^{3+1}$ invariant under P and CP transformations.
Moreover, it is contended that $f_{QCD}\left(  a/\phi\right)  $ also exhibits
P and CP odd behavior, contingent upon the existence of the axion field
$a\left(  x\right)  $ associated with the Peccei-Quinn symmetry in the strong
interactions \cite{PQ}. In such an instance, $S_{\text{$CS-QCD$}}^{3+1}$
retains P and CP invariance, akin to $S_{\text{$CS-QED$}}^{3+1}$. However, in
the event of the axion's absence, theoretical assumptions regarding the P and
CP properties of $f_{QCD}${} become uncertain. It could potentially function
as a dimensionless quantity dependent on all $s_{i}/\phi$, notwithstanding
experimental observations indicating its non-existence or extreme rarity,
leaving the underlying reasons for its minute magnitude unresolved, akin to
other unexplained hierarchies. Lastly, the function $f_{GR}\left(  s_{i}%
/\phi\right)  $ remains unconstrained by either theoretical postulations or
experimental evidence, rendering its P and CP properties undetermined.

In more detail, the action $S_{\text{$CS-QED$}}^{3+1}$ is not part of the
fundamental renormalizable action, but it arises in quantum loop corrections
and is sometimes included in effective actions involving certain hadrons that
decay into two photons. The coefficient $f_{QED}$ that arises from the chiral
triangle anomaly was computed reliably in perturbative QCD, and the result
agrees quantitatively with the measurement of the decay of the neutral pion
into two photons. Therefore, we conclude that $f_{QED}(\pi^{0}/\phi)$ in the
Weyl invariant i(SM+GR)$_{3+1},$ when taken at low energies in the c-gauge
$\phi\left(  x\right)  \rightarrow\phi_{0},$ is already known in the form
$f_{QED}(\pi^{0}/\phi_{0})=\left(  c/\phi_{0}\right)  \pi^{0}\left(  x\right)
$ where the dimensionful constant $\left(  c/\phi_{0}\right)  $ is given by
$\alpha/\left(  16\pi f_{\pi}\right)  $ in terms of measured quantities:
namely, the dimensionless $\alpha=1/137$ is the electromagnetic fine structure
constant, and the dimensionful $f_{\pi}$ is the pion decay constant. This
fixes the unknown dimensionless coefficient $c.$ Therefore, before fixing the
Weyl gauge we can write the fully determined dimensionless and scale invariant
$f_{QED}$ as
\begin{equation}
f_{QED}(\frac{\pi^{0}}{\phi})=\alpha\frac{\phi_{0}}{16\pi f_{\pi}}\frac
{\pi^{0}\left(  x\right)  }{\phi\left(  x\right)  }.\label{fQED}%
\end{equation}
In the c-gauge, we observe that $\phi_{0}/\phi_{c}\left(  x\right)  =1$,
yielding $f_{QED}=\alpha\frac{1}{16\pi f_{\pi}}\pi_{c}^{0}\left(  x\right)  $
wherein $\pi_{c}^{0}\left(  x\right)  $ represents the measured and
interpreted pion field at low energies. Significantly, the gauge-invariant
form outlined in (\ref{fQED}) proves instrumental in bridging the low-energy
description within the c-gauge to all regions encompassed by the geodesically
complete theory, facilitated by the Weyl gauge-invariant ratios $\pi
^{0}\left(  x\right)  /\phi\left(  x\right)  =\pi_{c}^{0}\left(  x\right)
/\phi_{0}.$ This correlation enables computations of $\pi^{0}\left(  x\right)
/\phi\left(  x\right)  $ in proximity to gravitational singularities to be
directly linked to the observed low-energy behavior.

In QCD, $S_{\text{$CS-QCD$}}^{3+1}$ is not part of the fundamental
renormalizable action unless $f_{QCD}$ is a constant. However, since it
violates P and CP in the strong interactions it is not introduced as a
fundamental constant. Nevertheless, $f_{QCD}$ arises in quantum loop
corrections when the electroweak interactions are coupled to quarks although
the non-perturbative nature of strong interactions renders the computation of
$f_{QCD}$ elusive, relegating it to an enigmatic effective term. In
i(SM+GR)$_{3+1}$ $f_{QCD}(s_{i}/\phi)$ must be a general dimensionless
function due to the Weyl symmetry. Given the presence of P\&CP violations in
the weak interactions that are part of the SM, it is hard to understand why
such $f_{QCD}$ must be unnaturally minute to reconcile the lack of P and CP
violation in experiments involving the neutron's electric dipole moment. This
discrepancy raises the fundamental issue of elucidating why $f_{QCD}${} should
be tuned to such diminutive proportions while $f_{QED}$ in QED remains non-negligible.

However, under the auspices of a Peccei-Quinn symmetry in the strong
interactions \cite{PQ},{} $f_{QCD}$ assumes linearity with respect to the
axion field $a\left(  x\right)  $. In the realm of Weyl symmetric theory, it
adopts the form $f_{QCD}\left(  a/\phi\right)  =c\times\left(  a\left(
x\right)  /\phi\left(  x\right)  \right)  $, where $\phi$ is symmetric under
P\&CP while $a(x)$ represents the pseudoscalar axion field, exhibiting oddness
under both P\&CP. This configuration yields the density $f_{QCD}\times\left(
\tilde{F}F\right)  _{QCD}$ that conserves both P\&CP symmetries. This
formulation resolves the strong CP problem by leveraging the PQ symmetry to
assert the vanishing vacuum expectation value of the axion, thereby ensuring
conservation of P\&CP in QCD, akin to QED's conservation involving the pion.

In the absence of PQ symmetry or its associated axion, the negligible size of
$f_{QCD}$ is possible, but it is unexplained like a few other unnatural
hierarchy problems in the Standard Model.

In the case of gravity, a more general function $f_{GR}\left(  s_{i}%
/\phi\right)  $ that is not purely odd or purely even under P and CP is
permitted within existing experimental constraints. Moreover, since GR is
already non-renormalizable, the action $S_{\text{$CS-GR$}}^{3+1}$ is not
prevented from appearing in the action like Eq.(\ref{ActioA2}) with its own
independent parameters, rather than being computed in quantum loops. If
$f_{GR}\left(  s_{i}/\phi\right)  $ is odd under P and CP, then the
$f_{GR}(s_{i}/\phi)\tilde{R}R$ CS density is symmetric under P and CP. On the
other hand, if there are only 2 spin 0 fields $\left(  \phi,s\right)  ,$
recalling that both $\phi$ and the Higgs $s$ are both P and CP even, then
$f_{GR}\left(  s/\phi\right)  $ is automatically even under both P and CP. In
that case, the CS term $f_{GR}\left(  s/\phi\right)  \,\tilde{R}R$ violates
both P and CP.

Future phenomenological studies are needed to constrain the exact forms of the
functions $f_{GR}\left(  s/\phi\right)  $, $f_{QCD}\left(  a/\phi\right)  $.
More generally, $f_{GR}\left(  s_{i}/\phi\right)  $ and $f_{QCD}\left(
s_{i}/\phi\right)  $ may involve several more fundamental spin-0 fields beyond
the Higgs (and the axion) if more fields are present in Nature.

\section{Discussion and outlook}

\label{sec:Discussion}

We have successfully modified the Chern-Simons terms in gravity, QCD, and QED
to accommodate local conformal scale (Weyl) invariance. Commencing with the
conformally improved i(SM+GR)$_{3+1}$ \cite{BST}, we extended this framework
by formulating Weyl symmetric CS terms.

First, we devised a Weyl invariant formulation of the Chern-Simons term in
gravity directly in 3+1 dimensions. We found that the local scale symmetry of
the i(SM+GR)$_{3+1}$ model could be upheld by linearly coupling the Pontryagin
density $\tilde{R}R$ in the Chern-Simons term to a function of the ratio of
conformally coupled scalar fields, $f_{GR}(s_{i}/\phi)$, as in
Eqs.(\ref{RRweyl},\ref{eq:CS-CC}). This form of $f_{GR}(s_{i}/\phi)$ proved
indispensable in achieving local conformal scale invariance. Similarly, we
discerned that the Chern-Simons terms in QCD and QED could attain local Weyl
symmetry if the density $\tilde{F}F$ linearly couples to a function
$f_{QCD}(s_{i}/\phi)$ or $f_{QED}(s_{i}/\phi)$ as in Eqs.(\ref{grQCDQED}).

The physical interpretation at low energies unfolds in the c-gauge, where the
i(SM+GR)$_{3+1}$ model virtually coincides with the customary SM+GR as
discussed in section \ref{DimParm}. In the c-gauge of (SM+GR+CS)$_{3+1}$, we
denote all fields with the letter `c' such as $\left(  \phi_{c}\left(
x\right)  ,s_{ci}\left(  x\right)  ,g_{c\mu\nu}\right)  $ to differentiate
them from other gauges. The c-gauge is delineated by the gauge choice
$\phi_{c}\left(  x\right)  =\phi_{0}$ where $\phi_{0}${} remains constant
across all $x^{\mu}$ within our spacetime patch as observers situated outside
all gravitational singularities. Within this spacetime patch, the c-gauge
scalars $s_{ci}\left(  x\right)  $ represent the spin 0 fields (e.g. the
Higgs, axion, etc.) as measured from our cosmological and accelerator physics
perspectives, conforming to their formulation in the standard SM+GR. The
invariance of field ratios under Weyl gauge transformations enables us to
express $s_{i}\left(  x\right)  /\phi\left(  x\right)  =s_{ci}\left(
x\right)  /\phi_{0}$. On the left hand side the fields are in any gauge, on
the right-hand side, we can measure both $s_{ci}\left(  x\right)  $ and
$\phi_{0}$ as explained in section \ref{DimParm}. This relationship
facilitates the linkage of the low-energy interpretation of the scalars
$s_{ci}\left(  x\right)  $ from the low-energy gravity patch to the
geodesically complete full spacetime in any gauge encompassing the vicinity of
every singularity and anti-gravity patches behind each singularity.

Thus, computations can be executed in any convenient gauge within the complete
spacetime of (SM+GR+CS)$_{3+1}$, and the gauge-invariant information
$s_{i}\left(  x\right)  /\phi\left(  x\right)  $ can be translated into the
low-energy language of the c-gauge (refer to the example in Eq.(\ref{fQED})).
Other gauge-invariant ratios $s_{i}/s_{j}$ can be re-expressed in terms of
$s_{i}/\phi$.

It is evident that at this stage, only the Weyl invariant Higgs field
$s/\phi=\sqrt{2H^{\dagger}H/\phi^{2}}$ is confirmed to exist in Nature. Thus,
we ascertain that at least two scalar fields $\left(  \phi,H\right)  $ are
constituents of the Weyl invariant theory (SM+GR+CS)$_{3+1}$. This theory
adeptly encapsulates all known facets of particle physics and gravity in any
gauge. This comprehensiveness stems from our understanding that in our
low-energy spacetime patch, the gauge-invariant $s\left(  x\right)
/\phi\left(  x\right)  \sim10^{-17}${} is minuscule, and in this limit, the
Weyl invariant theory closely mirrors the remarkably accurate conventional SM+GR.

In section \ref{sec:2T-physics-approach}, we determined the 4+2 dimensional
counterparts for the 3+1 CS terms. Recognizing that the locally conformally
invariant i(SM+GR)$_{3+1}$ serves as a holographic image of a 4+2 dimensional
field theoretical action, we delved into the 4+2 dimensional counterparts of
the 3+1 CS terms within the 2T-physics formalism. Successfully formulating the
actions $S_{\text{$CS-GR$}}^{4+2}${} for gravity, QCD, and QED, we
demonstrated how the emergent 3+1 actions $S_{\text{$CS-GR$}}^{3+1}${} are
derived as a holographic shadow of 2T-physics.

In the process, we clarified how Weyl symmetry in 3+1 dimensions in the
complete action (SM+GR+CS)$_{3+1}$ arises from a more fundamental general
coordinate invariance in the 4+2 dimensional parent theory (SM+GR+CS)$_{4+2}$.
It is crucial to note that Weyl symmetry was not among the gauge symmetries of
the 4+2 dimensional $S_{\text{SM+GR+CS}}^{4+2}${} actions; rather, it emerged
in the 3+1 holographic conformal shadow from general coordinate
transformations in 4+2 dimensions that intertwine the apparent 3+1 dimensions
with the concealed 1+1 large (not curled up) dimensions.

Thus, the genesis of Weyl symmetry within our work bears the hallmark of 2T
physics. While reminiscent of the original Weyl transformations, it diverges
conceptually since Weyl's concepts bore no relation to the additional hidden
dimensions pertinent to 4+2 spacetime. Furthermore, Weyl's geometry
\cite{Geometry} included a physical vector gauge field degree of freedom that
is absent in our scenario.

The incorporation of local scale symmetry renders our improved
(SM+GR+CS)$_{3+1}$ action geodesically complete by incorporating additional
spacetime behind gravitational singularities. This completeness is valid in
the presence of Chern-Simons corrections to GR, QCD, and QED. The symmetric
formalism furnishes analytical control through singularities and steers
physical interpretation by tethering it to the low-energy interpretation of
the fields. New physics manifests as the factor $\left(  \phi^{2}\left(
x\right)  -s^{2}\left(  x\right)  \right)  $ multiplying curvature in the
action (\ref{ActioA2}) can vanish and change sign in regions of spacetime
where the gauge-invariant Higgs field $\sqrt{2H^{\dagger}H}/\left\vert
\phi\right\vert $ burgeons to approach the value 1.

By examining the equations of motion, it becomes apparent that spacetime
singularities manifest precisely when $\sqrt{2H^{\dagger}H}/\left\vert
\phi\right\vert =1.$ In regions of spacetime where singularities may arise,
strong gravity dominates, driven by the remarkable behavior of the effective
spacetime-dependent gravitational strength $G\left(  x\right)  $ satisfying
$16\pi G\left(  x\right)  =12\left(  \phi^{2}\left(  x\right)  -s^{2}\left(
x\right)  \right)  ^{-1}$ which diverges. It is noteworthy that in proximity
to such singularities, the magnitude of $\sqrt{2H^{\dagger}H}/\left\vert
\phi\right\vert $ can approach unity even if the gauge-dependent fields $\phi$
and $H$ are both either individually small, individually large, or
intermediate. In previous studies in a cosmological setting, it was revealed
that at the singularity both $\phi$ and $H$ remarkably vanished at the Big
Bang singularity where $\sqrt{2H^{\dagger}H}/\left\vert \phi\right\vert =1$
\cite{BbBc}$,$ indicating that the electroweak SU$\left(  2\right)  \times
$U$\left(  1\right)  $ symmetry is restored at the singularity. This and
similar manifestations of new physics are completely unexpected in the
conventional SM+GR.

Extending the leads of (SM+GR+CS)$_{3+1}$ and (SM+GR+CS)$_{4+2}${} toward an
enhanced version of string theory that is capable of reproducing these Weyl
symmetric and geodesically complete field theories at their low-energy limits
stands as a promising avenue for future research. An essential requirement for
such an improved string theory is the emergence of the dimensionful string
tension from a field, mirroring how all dimensionful parameters arise from
$\phi$ within the field theory framework. Progress in this direction was
initiated in \cite{BSTstrings}. Accomplishing this objective would lay the
groundwork for establishing a coherent quantum treatment of these innovative
theories that ensures geodesic completeness.

Another intriguing avenue for investigation involves delving into the realms
of field theory beyond the conformal shadows discussed in this study. This
exploration not only facilitates the development of duality relationships
among shadows but also offers avenues for crafting new computational tools
within 1T-physics at a fundamental field theoretic level. It is conceivable
that frameworks such as the AdS/CFT duality could find elucidation through
this approach alongside the potential discovery of novel dualities. Numerous
striking examples of multi-dualities already exist in classical and quantum
physics as exemplified by \cite{Survey}-\cite{IB+Araya}. The objective here is
to develop analogous methodologies in the context of field theory for shadows
derived from (SM+GR+CS)$_{4+2}$. Preliminary explorations of such field theory
dualities and hidden symmetries related to the 4+2 dimensions are illustrated
with a few simple examples in \cite{DualFieldThs}. These endeavors not only
shed light on the existence and nature of the additional 1+1 large dimensions
but also underscore the significance of constructing the parent theory in 4+2
dimensions as demonstrated in this paper for (SM+GR+CS)$_{4+2}$. Future
investigations in this domain are bound to unveil deeper insights into the
fabric of spacetime.

\appendix

\section{An overview of 2T-physics \label{A1}}

2T-physics originated in 1998 from a fundamental gauge symmetry, Sp($2,R$),
manifest in the phase space $\left(  X^{M},P_{M}\right)  $ within a 4+2
dimensional framework (more generally, $d+2$) \cite{Survey}-\cite{book}. The
basic notion is the postulate that the fundamental rules of physics should be
invariant under phase space transformations that generalize Einstein's general
coordinate transformations. This was accomplished with the Sp($2,R$) gauge
symmetry that encapsulates three generators that are functions of phase space.
These are represented as symmetric tensors $Q_{ij}\left(  X,P\right)  $ where
$i=1,2$, and they conform to the closure condition under Poisson brackets,
forming the Sp($2,R$) Lie algebra. A plethora of such structures exists with
an infinite variety of configurations \cite{IBDeli}\cite{U1}. Among them, the
simplest set, delineated below, serves as a foundational example:
\begin{equation}
Q_{11}=\frac{1}{2}X\cdot X,\;\;Q_{12}=\frac{1}{2}X\cdot P=Q_{21}%
,\;\;Q_{22}=\frac{1}{2}P\cdot P. \label{X2P2}%
\end{equation}
The dot product in these elemental expressions can possess an arbitrary
signature within any flat geometry, without impacting the Lie algebra.
However, for physical viability, only the 4+2 signature (or more generally,
$d+2$ with $d\geq2$) remains physically pertinent, as elucidated below.

\subsection{2T-physics in the phase space of a single particle
\label{1particle}}

Consider the action governing the worldline dynamics of a single spinless
particle, imbued with Sp$(2,R)$ gauge invariance in phase space:
\begin{equation}
L=\dot{X}\left(  \tau\right)  \cdot P\left(  \tau\right)  -\frac{1}{2}%
A^{ij}\left(  \tau\right)  Q_{ij}\left(  X\left(  \tau\right)  ,P\left(
\tau\right)  \right)  ,
\end{equation}
where $A^{ij}\left(  \tau\right)  $ represents the 2$\times2$ symmetric
Yang-Mills type Sp$\left(  2,R\right)  $ gauge fields on the worldline and
$Q_{ij}\left(  X,P\right)  $ denotes the 2$\times2$ symmetric generators of
Sp$(2,R)$ transformations acting on the "matter" variables $\left(
X^{M},P_{M}\right)  ${} \cite{Survey}-\cite{IB+Araya}.

Turning our attention to the gauge-invariant physical sector of the theory,
corresponding to Sp(2,R) singlets invariant under Sp(2,R) transformations, we
impose the condition that all generators of Sp(2,R) vanish: $Q_{ij}\left(
X,P\right)  =0$. This requirement, enforced by the equation of motion for
$A^{ij}$, necessitates specific conditions on the phase space, exemplified by
the simple form of $Q_{ij}$ in Eq.(\ref{X2P2}), $X^{2}=P^{2}=X\cdot P=0$.
Importantly, for nontrivial solutions to exist, the flat metric in these dot
products, $\eta_{MN}$, must possess two or more timelike
directions.\footnote{For 0 times, the solutions are $X^{M}=P^{M}=0.$ For 1
time, both $X^{M}$ and $P^{M}$ must be lightlike and parallel to each other,
hence there is no angular momentum. Therefore, both cases describe trivial
motions. With 2 or more times, there are an infinite number of solutions.}
However, if timelike directions exceed two, the gauge symmetry becomes
insufficient for eliminating negative probability ghosts. Thus, to ensure
nontrivial and ghost-free physical solutions that remain Sp(2,R)
gauge-invariant, while adhering to unitary and causal scenarios, the phase
space---including gauge degrees of freedom---must contain precisely two
timelike dimensions---no more and no less \cite{GMann}].

Compared with 1T-physics, which typically involves one constraint (e.g.,
$p^{2}+m^{2}=0$ for a freely moving relativistic particle, or
generalizations), 2T-physics imposes two additional constraints. This
augmented gauge symmetry allows removal of one timelike and one spacelike
dimension from the phase space via gauge fixing, and resolving two out of the
three constraints. Consequently, 2T-physics encompasses one extra timelike and
one extra spatial dimension compared to 1T-physics. The gauge-invariant
physical sector in 4+2 dimensions (more generally $d+2$) resembles 1T-physics
in 3+1 dimensions (more generally $\left(  d-1\right)  +1$).

In this formalism of theoretical physics, intriguing connections emerge
between seemingly disparate phenomena. Among these connections are the
shadows---alternate descriptions or formulations---of physical systems with
different Hamiltonians as described below.

The solutions satisfying the simplest constraints (such as Eq. (\ref{X2P2}))
are termed holographic shadows. These shadows encapsulate all 1T-physics
systems describable by Hamiltonians derived from a single particle's phase
space. Thus, all 1T-physics manifestations are unified into the trio of
2T-physics equations, $X^{2}=P^{2}=X\cdot P=0$, highlighting a profound
unification of all 1T-physics Hamiltonian systems as emerging only from the
Sp$\left(  2,R\right)  $ constraints.

Shadow actions, representing gauge-fixed solutions of Eq.(\ref{X2P2}),
delineate 1T-physics systems with varied Hamiltonians within the 1T formalism.
The profusion of shadows arises from the myriad ways of embedding 3+1
dimensional phase space $\left(  x^{\mu},p_{\mu}\right)  $ into 4+2
dimensional phase space $\left(  X^{M},P_{M}\right)  $, leading to diverse
gauge-fixing methods and resolutions of Sp$(2,R)$ constraints. Consequently,
multiple manifestations of the same 4+2 phenomena emerge as shadows in the
parlance of 1T-physics. Remarkably, besides the three constraints $X^{2}%
=P^{2}=X\cdot P=0$, no additional equations are required to capture and unify
all single-particle 1T-physics dynamics for a single spinless particle. This
encompasses all Hamiltonians, incorporating any associated parameters such as
mass and interaction parameters, which emerge as moduli within the diverse
embeddings of 3+1 dimensional phase space into 4+2 dimensional phase space.

Evidently, due to the dot product's SO$(d,2)$ invariance, all shadows possess
global SO$(d,2)$ spacetime symmetry, with conserved generators represented by
$L^{MN}=\left(  X^{M}P^{N}-X^{N}P^{M}\right)  .$ SO$\left(  d,2\right)  $
transformation rules for the full, overt, or covert SO$(d,2)$ symmetry in each
shadow action are generated by the SO$(d,2)$ generators $L^{MN}$ where
$\left(  X^{M},P^{N}\right)  $ are replaced by their gauge-fixed versions,
$X^{M}\left(  x^{\mu},p_{\mu}\right)  $ and $P^{N}\left(  x^{\mu},p_{\mu
}\right)  $, expressed in 3+1 dimensional phase space $\left(  x^{\mu},p_{\mu
}\right)  $ of the given shadow.

The gauge group Sp$\left(  2,R\right)  $ is generalized into a supergroup with
the inclusion of extensions to phase space accommodating spin \cite{Gauged}
and/or supersymmetry \cite{SUSY2T}. In the realm of local field theory, local
fields in 2T-physics---be they scalars, vectors, tensors, or spinors---extend
not only as functions of a larger spacetime $X^{M}$ but also encompass tensor
or spinor components along the additional 1+1 directions. The momentum $P_{M}%
${} transitions into covariant derivatives acting upon such fields.

These shadows exhibit multi-dualities under Sp$(2,R)$ gauge transformations,
transitioning between fixed gauges. For instance, a spinless particle that
obeys the Sp$\left(  2,R\right)  $ constraints in a flat d+2 dimensional
spacetime, manifests as various shadows in 3+1 dimensions, including but not
limited to those in the following list \cite{Survey}-\cite{IBQuelin}: a free
massless relativistic particle, a free massive relativistic particle, a free
massive non-relativistic particle, a particle in anti-de-Sitter space
AdS$_{d}$, a particle in de-Sitter space dS$_{d}$, a particle in any maximally
symmetric space (e.g., AdS$_{d-n}\times$S$^{n}$), a particle in the
Robertson-Walker spacetime (for both open and closed universes), BTZ black
holes (for $d$=$3$), the non-relativistic Hydrogen atom, the non-relativistic
harmonic oscillator, amongst others, and a twistor reformulation of all of
these systems.

Remarkably, each 3+1 shadow fully encapsulates holographically the Sp$(2,R)$
gauge-invariant physical essence of the parent 4+2 dimensional theory.
Consequently, all shadows exhibit gauge equivalence under Sp$(2,R)$ gauge
transformations, taking the form of canonical transformations within
1T-physics that include transformations of time, Hamiltonian, and three
positions and momenta. These canonical transformations among shadows are
construed as multi-duality transformations within the framework of 1T-physics
\cite{IB+Araya}. Thus, 2T-physics not only unveils hidden extra 1+1 large
dimensions, treated equivalently to the overt 3+1 dimensions but also unveils
numerous unsuspected multi-dualities, serving as novel tools for duality-based
computations within 1T-physics while deepening understanding of spacetime's
fundamental nature.

Despite their diverse expressions as Hamiltonians in phase space in $(d-1)+1$
dimensions, all shadow actions share a fundamental underlying hidden spacetime
symmetry in $d+2$ dimensions, revealing the extra $1+1$ dimensions.
Specifically, shadows arising from $d+2$ dimensional flat spacetime encompass
a vast array of $(d-1)+1$ dimensional spacetimes. At the quantum level, all
emerging 1T dynamical shadow systems represent physically distinct 1T
manifestations of the same unifying mathematical unitary representation of
SO$(d,2)$ common to all shadows. This unique representation of SO$\left(
d,2\right)  ,$ known as the singleton representation, emerges directly from
the covariant quantization treatment of the 2T system in $d+2$ dimensions,
consisting of the vanishing of 3 three hermitian generators of Sp$\left(
2,R\right)  ,$ namely $X^{2}=P^{2}=\left(  X\cdot P+P\cdot X\right)  =0.$ This
singleton representation is distinguished solely by its unique Casimir
eigenvalues which are determined exclusively by the number of dimensions $d$
(see Eq.9 in \cite{Oscillator}). In essence, the Hilbert spaces for the
$(d-1)+1$ dimensional shadows are expressed as unitarily equivalent bases of
the same singleton representation of SO$(d,2)$. Across various 1T shadows,
their quantum Hilbert bases differ only by the choice of a subset of
simultaneous observables, all of which are functions of the generators of
SO$\left(  d,2\right)  $, namely $L^{MN}=X^{M}P^{N}-X^{N}P^{M},$ while the
Casimir eigenvalues remain unaltered for all shadows. Remarkably, this
prediction is corroborated by computations directly in 1T-physics, where the
quantum spectra of systems such as the hydrogen atom, harmonic oscillator, and
others confirm the presence of the hidden SO$(d,2)$ symmetry, with the
anticipated eigenvalues of the Casimir operators \cite{Oscillator}%
\cite{Symmetries}.

Of particular importance is the \textquotedblleft conformal
shadow,\textquotedblright\ where the embedding of 3+1 phase space into 4+2
phase space aligns with the formalism of relativistic 3+1 spacetime,
encompassing conformal symmetry SO$(4,2)$. In this shadow, the partially
concealed nonlinear special conformal transformations in 3+1 spacetime
$x^{\mu}$ emerge from the explicit linear Lorentz transformations SO$(4,2)$ in
4+2 spacetime $X^{M}$, including the extra 1+1 spacetime dimensions as gauge
degrees of freedom \cite{Survey,GMann,IB+Araya}. The significance of the
conformal shadow lies in its direct correlation with relativistic field theory
in 1T-physics. Through this connection, 2T-physics stands as a robust
descriptor of nature across all energy or distance scales known to date, while
predicting new hidden symmetries and dualities discernible with tools of 1T-physics.

According to 2T-physics, all 1T-physics systems have hidden symmetries that
relate to the presence of extra 1+1 dimensions. The key to revealing the
hidden spacetime structure of a given 1T-physics system lies in constructing
the 4+2 dimensional parent that underlies its evident 3+1 spacetime structure
(not necessarily expressed in relativistically covariant notation). Viewing
the latter as a shadow of 2T-physics opens up a pathway to unification. The
parent structure predicts multi-dual shadows, allowing us to unify disparate
1T systems. In doing so, we expand our understanding of spacetime and gain a
powerful tool for predicting dualities and hidden symmetries within 1T-physics.

This pursuit of unification and discovery lies at the heart of 2T-physics. In
this spirit, we delved into the 3+1 Chern-Simons action, denoted as
$S_{\text{$CS-GR$}}^{3+1}$, in section (\ref{sec:building-weyl-symmetric}) and
followed up by constructing the parent action $S_{\text{$CS-GR$}}^{4+2}$ in
section (\ref{sec:2T-physics-approach}). In turn, starting with the parent
$S_{SM+GR+\text{$CS$}}^{4+2}$ it is in principle possible to derive various
holographic shadows in the context of field theory, beyond the conformal
shadow, study the multi-dualities among them for their inherent intereting
properties and hopefully develop useful new tools for computations based on dualities.

\subsection{Field Theory in 2T-Physics \label{2Tfields}}

In the framework of 2T field theory formalism, the architecture of the 2T
action is designed to enforce the vanishing of Sp$(2,R)$ generators (and their
extensions) through equations of motion derived from the 2T action principle
(refer to section \ref{sec:2T-physics-approach} and Appendix \ref{A2}). This
2T-physics field theory encompasses interactions ranging from Yang-Mills gauge
symmetry in d+2 dimensions to general coordinate reparametrization symmetry
and even supergravity in d+2 dimensions.

When exploring classical or quantum mechanics within the 4+2 dimensional
theory, as well as in the context of field theory, a myriad of 3+1 dimensional
holographic representations can be obtained, as discussed in the previous
section, by gauge fixing and solving two (out of three) of the Sp$(2,R)$
constraint equations of motion. In the field theory version, these are termed
\textquotedblleft kinematic equations\textquotedblright\ as discussed in
Appendix \ref{A2}. These kinematic constraints are independent of field
interactions, while the remaining third \textquotedblleft dynamical
constraint\textquotedblright\ encompasses the mentioned interactions.
Essentially, the solution space of the kinematic constraints suggests that for
a given action in 2T-physics field theory in 4+2 dimensions, numerous emergent
field theory actions (referred to as shadows) in 3+1 dimensions are in
principle feasible.

Among these diverse shadows, particular attention is drawn to the
\textit{holographic conformal} shadow in the current discourse. This shadow
conspicuously exhibits the attributes of special or General Relativity,
serving as the conduit from 2T-physics to relativistic field theory in
1T-physics. It is within this shadow that all relativistic field theories
manifest, including the empirically successful Standard Model (SM)
\cite{2Tsm}, General Relativity (GR) \cite{2TgravIB,2Tgrav}, and their
supersymmetric extensions. The coupled system in 1T-physics i(SM+GR)$_{3+1}$
\cite{BST} is deemed the conformal shadow of the parent field theory
(SM+GR)$_{4+2}$ as elaborated in section \ref{sec:2T-physics-approach}. It is
noteworthy that the parent (SM+GR)$_{4+2}$ theoretically yields a multitude of
shadows, including the conformal shadow i(SM+GR)$_{3+1}$, which are multi-dual
field theories to one another. Various simple examples of such dual shadows
within the realm of field theory were expounded upon in \cite{DualFieldThs}%
\cite{IBQuelin}. This aspect of field theoretic duality within 2T-physics
stands open to broader exploration, offering avenues for the development of
novel duality-based computational methodologies.

\subsection{Sp$\left(  2,R\right)  $ gauge invariant sector and holographic
reduction of 4+2 to 3+1 \label{A2}}

In this section, we elucidate the significance of the field $W(X)$ introduced
in section (\ref{sec:2T-physics-approach}) in the delta function
$\delta(W\left(  X\right)  )$. As discussed in \cite{2Tsm,2TgravIB,2Tgrav},
the presence of $W(X)$ and its inclusion in the delta function stem from the
foundational formulation of 2T physics, which hinges on a broader gauge
symmetry in phase space $\left(  X^{M},P_{M}\right)  $ rather than solely
position space $X^{M}$. In a flat space scenario, $W\left(  X\right)  $ is
denoted by $W_{\text{flat}}\equiv\left(  X^{2}\right)  _{\text{flat}}%
=X^{M}X^{N}\eta_{MN}$ where $\eta_{MN}$ represents a flat metric in 4+2
dimensions. This $X^{2}$ stands as one of the generators of the fundamental
Sp$(2,R)$ gauge symmetry in flat phase space as displayed in Eq.(\ref{X2P2}).
The other two generators, in the absence of interactions in flat space and
after quantum ordering, are $\frac{1}{2}\left(  X\cdot P+P\cdot X\right)
_{\text{flat}}$ and $\left(  P^{2}\right)  _{\text{flat}}$. These three
quadratic phase space quantities collectively form the Sp$(2,R)$ Lie algebra,
closing under both Poisson brackets and quantum commutators.

In interacting field theory within curved space, $\left(  X^{2}\right)
_{\text{flat}}$ is elevated to the field $W(X)$, with momenta transformed into
derivatives $P_{M}\rightarrow-i\partial_{M}$, and interactions integrated
across all fields, incorporating a metric $G_{MN}\left(  X\right)  $ in curved
space. Within the gauge invariant sector of 2T physics, known as the singlet
sector of Sp$(2,R)$ and deemed physical, all three generators of Sp$(2,R)$
must vanish. The delta function $\delta(W\left(  X\right)  )$ imposes the
vanishing of one of these three Sp$(2,R)$ gauge symmetry generators, $W(X)=0$,
thus partially enforcing the underlying phase space Sp$(2,R)$ gauge invariance
and consequently eliminating part of the gauge degrees of freedom. This
process selectively emphasizes Sp$(2,R)$ gauge invariants while minimizing
gauge degrees of freedom. It is essential to note that the other two
generators of Sp$(2,R)$ must also vanish to fully concentrate solely on the
gauge invariant physical degrees of freedom.

We enforce the vanishing of these generators by solving specific constraint
equations of motion derived from varying the 2T field theory action which
incorporates the delta function $\delta(W\left(  X\right)  )$. For instance,
upon varying the action in Eq.(\ref{eq:fullS}) for each field (e.g., the
$\Phi$ field), a linear combination of the delta function and its derivatives
is obtained \cite{2TgravIB},
\begin{equation}
A_{\Phi}\delta(W)+B_{\Phi}\delta^{\prime}(W)+C_{\Phi}\delta^{\prime\prime
}(W)=0.
\end{equation}
Given that $\delta(W),\delta^{\prime}(W)$, and $\delta^{\prime\prime}(W)$ are
linearly independent, this yields three equations, $\left.  A_{\Phi
}\right\vert _{W=0}=\left.  B_{\Phi}\right\vert _{W=0}=\left.  C_{\Phi
}\right\vert _{W=0}=0$, and similarly for all fields $\Phi,S_{i},\Psi_{\alpha
},A_{M}^{a},G_{MN}$ in addition to those displayed for $\Phi.$ These three
differential equations (and associated equations for tensors and spinors
restricting their directions) enforce the vanishing of all Sp$(2,R)$
generators when applied to the fields $G_{MN},\Phi,S_{i},\Psi_{\alpha}%
,A_{M}^{a}$ \cite{2Tsm}-\cite{2Tgrav}. The solutions to these equations yield
the physical Sp$(2,R)$ gauge invariant sector of the theory. The B and C
conditions (and their tensorial and spinorial counterparts) are termed the
\textquotedblleft kinematic equations," while the A condition is labeled the
\textquotedblleft dynamical equation." Notably, only the A equation depends on
interactions. The solutions of the kinematic equations, independent of
interactions, yield the 3+1 dimensional \textquotedblleft shadows" for each
field and their \textquotedblleft prolongations" into 4+2 dimensions. The
shadow represents the analog of the bottom mode $\left(  g_{\mu\nu},\phi
,s_{i},\psi,A_{\mu}^{a}\right)  $ in a Kaluza-Klein type expansion, whereas
the prolongations are conceptually akin to higher Kaluza-Klein type modes as
illustrated in the series expansions Eq.(\ref{expand}).

However, unlike the conventional Kaluza-Klein case, within the framework of
2T-physics, the A, B, C equations restrict the prolongations, rendering them
not independent modes. Instead, the prolongations are entirely determined by
the shadows of each field $\left(  g_{\mu\nu},\phi,s_{i},\psi,A_{\mu}%
^{a}\right)  $ in 3+1 dimensions - except for unphysical unfixed prolongation
gauge degrees of freedom independent of the shadow fields. Thus, the Sp$(2,R)$
invariant physical sector emerges as a 3+1 dimensional field theory directly
treatable within 3+1 dimensions.

Applying this approach to the (SM+GR)$_{4+2}$ action presented in
Eq.(\ref{eq:fullS}) yields the conformal shadow i(SM+GR)$_{3+1}$ outlined in
Eq.(\ref{SM-GR}). This emergent 3+1 dimensional theory differs from the
conventional formulation of the Standard Model+General Relativity (SM+GR) in
several key aspects \cite{2Tsm}-\cite{2Tgrav}. Notably, it must adhere to
local scale invariance, with a requisite relative sign between the $\phi$ and
$s_{i}$ fields \cite{BST}, as depicted in Eq.(\ref{SM-GR}). These additional
properties are necessitated by the underlying Sp$(2,R)$ gauge symmetry,
including the local scale (Weyl) symmetry emerging from general coordinate
reparametrization that amalgamates the extra 1+1 dimensions with the 3+1
dimensions (refer to Eqs.(\ref{rescaleW},\ref{leftover})). This necessitates
that all scalar fields in relativistic field theory, including the Higgs, must
adhere to the principles of conformal coupling. If there exists more than one
s-type scalar, the Weyl symmetry can manifest in a broader and more nonlinear
fashion \cite{BST}.

The mandated Weyl symmetry illustrates how 2T-physics constrains the behavior
of the Higgs in this new emergent action, diverging in certain aspects from
the conventional Standard Model plus gravity paradigm. Although imperceptible
at low energies in particle accelerators, this deviation becomes significantly
pronounced in regions of spacetime characterized by strong gravity, such as
the interiors of black holes, where the physics of i(SM+GR)$_{3+1}$ diverges
markedly from the predictions of the standard theory \cite{BHoles}%
\cite{BH+Higgs}.

While our discussion in this paper primarily revolves around the conformal
shadow, it is noteworthy that other shadows stemming from the same 4+2 action,
contingent upon alternative 3+1 embeddings in 4+2 dimensions, harbor distinct
interpretations within 1T-physics. For some simple examples of other shadows
in the realm of field theory, refer to \cite{DualFieldThs}\cite{IBQuelin}. In
essence, these shadows represent specialized instances of potential 3+1
dimensional field theories that are mutually dual. The exploration and
development of such multi-dualities in field theory, alongside their
application in deriving novel computational methodologies within 1T-physics,
remain vibrant areas of investigation within the domain of 2T-physics.

\end{document}